\documentclass[apjl]{emulateapj}

\shorttitle{QUEST--La Silla AGN Survey}
\shortauthors{Cartier et~al.}

\begin{document}

\title{\bf The QUEST-La Silla AGN Variability Survey}
\author{R\'egis Cartier\altaffilmark{1, 2, 3}, Paulina Lira\altaffilmark{2}, Paolo Coppi\altaffilmark{4}, Paula S\'anchez\altaffilmark{2}, Patricia Ar\'evalo\altaffilmark{5},
Franz~E. Bauer\altaffilmark{1,6}, David Rabinowitz\altaffilmark{4}, Robert Zinn\altaffilmark{4}, Ricardo~R. Mu\~noz\altaffilmark{2} and Nicol\'as Meza\altaffilmark{2}}
\altaffiltext{1}{Millennium Institute of Astrophysics}
\altaffiltext{2}{Departamento de Astronom\'ia, Universidad de Chile, Casilla 36-D, Santiago, Chile}
\altaffiltext{3}{School of Physics \& Astronomy, University of Southampton, Southampton, Hampshire, SO17 1BJ, UK}
\altaffiltext{4}{Department of Astronomy, Yale University, New Haven, CT 06520-8101}
\altaffiltext{5}{Instituto de F\'isica y Astronom\'ia, Facultad de Ciencias, Universidad de Valpara\'iso, Gran Breta\~na 1111, Valpara\'iso, Chile}
\altaffiltext{6}{Instituto de Astrof\'isica, Pontif\'icia Universidad Cat\'olica de Chile, Casilla 306, Santiago, Chile}

\begin{abstract}

\noindent

We present the characterization and initial results from the QUEST-La Silla AGN variability survey.
This is an effort to obtain well sampled optical light curves in extragalactic fields with unique
multi-wavelength observations. We present photometry obtained from 2010 to 2012 in the XMM-COSMOS field,
which was observed over $150$ nights using the QUEST camera on the ESO-Schmidt telescope.
The survey uses a broadband filter, the $Q$-band, similar to the union of the $g$ and the $r$
filters, achieving an intrinsic photometric dispersion of $0.05$ mag, and a systematic error of $0.05$
mag in the zero-point. Since some detectors of the camera show significant non-linearity, we use
a linear correlation to fit the zero-points as a function of the instrumental magnitudes, thus obtaining
a good correction to the non-linear behavior of these detectors. We obtain good photometry to an equivalent
limiting magnitude of $r\sim 20.5$. The astrometry has an internal precision of $\sim 0\farcs1$ and an
overall accuracy of $0\farcs2$ when compared to SDSS. Studying the optical variability of X-ray
detected sources in the XMM-COSMOS field, we find that the survey is $\sim75-80$\% complete to magnitudes 
$r\sim20$, and $\sim67$\% complete to a magnitude $r\sim21$. Additionally, broad-line (BL) AGN have larger variation
amplitude that non-broad line (NL) AGN, with $\sim80$\% of the BL AGN classified as variable, whilst only
$\sim21$\% of the NL AGN resulted classified as variable. We also find that $\sim22$\% of objects classified 
as galaxies (GALs), are also variable. The determination and parameterization of the structure function 
($SF_{norm}(\tau) = A \tau^{\gamma}$) of the variable sources shows that most BL AGN are characterized by $A > 0.1$ and 
$\gamma > 0.025$. It is further shown that NL AGN and GAL sources occupying the same parameter space in $A$ and
$\gamma$ are very likely to correspond to obscured or low luminosity AGN. Our samples are, however, small, and
we expect to revisit these results using larger samples with longer light curves obtained as part of our ongoing survey.

\end{abstract}


\section{INTRODUCTION}

Despite the fact that variability is one of the defining characteristics of Active Galactic Nuclei (AGNs) we do not completely 
understand the mechanisms that drive such variation. Our understanding misses significant details of how AGN variability at different 
wavelengths is related\citep[see e.g.,][]{arevalo08, arevalo09, lira11, lira15}, and how physical properties of the central engine
(e.g., luminosity, black hole mass, hardness ratio, optical colors, etc) are related to well defined variability properties of the system
(e.g., characteristic timescale, variability amplitude, etc). From a cosmological perspective, the strong evolution of AGNs opens
the possibility to observe changes of the structure and feeding of AGNs over cosmic time \citep[see][and references therein]{shemmer14}.
In fact, controversy remains on the redshift dependence of AGN variability since the observed wavelength and the minimum luminosity
in a sample are highly correlated with redshift \citep[see][, and references therein]{vandenberk04, kelly09, macleod10, morganson14}.
Finally, the property of being highly variable makes AGN selection by variability a very promising tool to find them.

In order to study in detail the variability properties of individual AGNs and to use this information to rule out different
variability models to probe a wide range of timescales is required. Hence, long and intense campaigns are crucial. Fortunately, in recent
years, surveys covering a significant part of the sky, revisiting the same regions on timescales from days to years, and containing
a large sample of serendipitous objects --blind surveys-- are now becoming available as predecessors of the Large Synoptic Survey
Telescope \citep[LSST;][]{ivezic08}. LSST will revisit each part of the of the southern sky ($\delta_{J2000}$ \textless $+34.5$)
approximately every three nights over 10 years, and will observe in six bands ($ugrizy$) to a limiting magnitude of $r \simeq 24.5$ mag. 
Among the LSST predecessors we have: Sloan Digital Sky Survey (SDSS--\citep{york00}) equatorial Stripe-82 which covers roughly 280 deg$^2$ 
in the $20^{h}:34^{m}$--$04^{h}:08^{m}$
right ascension range, and declination \textbar$\delta_{J2000}$\textbar \textless $1.266^{\circ}$ \citep{ivezic04, sesar07}.
This region was observed in five bands ($ugriz$) once to three times a year from 2000 to 2005 (SDSS-I), and then from 2005 to
2008 at an increased cadence of 10 to 20 times per year as part of the SDSS-II supernova survey \citep{friedman08}.
The Catalina Real-Time Transient Survey \citep[CRTS;][]{drake09, graham14} which
covers between $-75^{\circ}$ \textless $\delta_{J2000}$ \textless $75^{\circ}$, observing in the $V$ filter to a limiting magnitude of $\sim 19$ to $21.5$
(depending on the telescope used). CRTS covers a total of $33,000$ deg$^{2}$ observing up to $\sim 2500$ deg$^{2}$ per night, with 4 exposures per visit,
separated by 10 min, over 21 nights per lunation (i.e., re-visiting a field every 10 to 15 days). The Palomar Transient Factory \citep[PTF;][]{rau09, law09}
covers between $-15^{\circ}$ \textless $\delta_{J2000}$ \textless $85^{\circ}$ using two broad-bandpasses, namely the Mould $R$ band and $g$ band,
to a limiting magnitude of $R \simeq 21.0$ mag and $g \simeq 21.6$ mag with a cadence of 3 to 5 days. The Panoramic Survey Telescope
and Rapid Response System \citep[PanSTARRS;][]{hodapp04, tonry12} covers all the northern sky down to $\delta \simeq -30^{\circ}$ in six bands 
\citep[$g_{p}$, $r_{p}$, $i_{p}$, $z_{p}$, $y_{p}$ and $w_{p}$;][]{tonry12} to a limiting magnitude between 22.0 mag to 24.0 mag depending on the filter,
re-visiting a field every 4 to 5 days. Finally, the Dark Energy Survey \citep[DES;][]{abbott05} is observing part of the southern sky in
four bands ($griz$) to a limiting magnitude of $r \simeq 24$ mag, performing a roughly 30~deg$^{2}$ time domain survey for supernovae 
re-visiting a field every few days (4 to 5 days). 

One of the advantages of the QUEST-La Silla AGN Survey over these surveys is the very intense monitoring, observing the survey fields
every possible night (although with large observing gaps from 2010 to 2012 -- see more details in Section \ref{fields} below).
We obtain between $2$ to $5$ observations per night to remove spurious variability due to artifacts, to potentially study intra-night
AGN variability, and to produce stacked images to reach deeper magnitudes. Individual images reach a limiting magnitude between
$r \sim 20.5$ mag and $r \sim 21.5$ mag for a exposure time of
60 seconds or 180 seconds, respectively. One of the main characteristics of the QUEST-La Silla AGN Survey is that is focused on
deeply observed extra-galactic fields with multi-wavelength coverage, and having nearly simultaneous observations in the near-infrared
(near-IR) performed by the Visible and Infrared Survey Telescope for Astronomy (VISTA) surveys (see Section \ref{fields} below).

X-ray variability studies indicate that the X-ray short-timescale normalization of the power spectrum density (PSD) is correlated
with the black hole mass \citep[see ][]{mchardy13, kelly13, gierlinski08, mchardy88}, and also that the black hole mass is correlated
with the break frequency (characteristic timescale) of the PSD \citep{mchardy06}. These results have lead to the suggestion that AGNs
are scaled versions of Galactic X-ray binaries with super-massive black holes \citep{mchardy06}, and that the X-ray characteristic
timescale and the black hole mass are tightly related. Thus AGN variability could be a robust black hole mass estimator for a
significant number of AGNs if these relations can be extended to the optical/ultraviolet (UV) wavelengths. 
For a long time a significant anticorrelation between AGN variability amplitude and luminosity has been recognized in the optical/UV regime
\citep{uomoto76, hook94, trevese94, cristiani97, vandenberk04, wilhite08, macleod10, meusinger13}. 
Besides, some correlation between characteristic variability timescales and black hole mass have already been found \citep{kelly09, collier01}. Additionally, correlations
between the optical variability amplitude and the black hole mass have been described \citep{wold07, wilhite08, kelly09, macleod10, meusinger13}.
However, these results seem to point towards a more fundamental inverse relation between the AGN variability amplitude, at timescales longer
than 1 year, and the Eddington ratio (or accretion rate) \citep{wilhite08, ai10, macleod10, zuo12, meusinger13}.
Further results indicate that AGN variability properties (e.g., the structure function) seem to change with other properties such as X-ray
luminosity, and radio loudness \citep[see ][ and references therein]{vandenberk04, macleod10}. All these results naturally lead to the
conclusion that AGN variability is intrinsically related to the physical parameters that govern the accretion in AGNs, and therefore,
variability studies using large samples are crucial to improve our understanding of the accretion process.

One of the reasons why these relations are not commonly used to estimate physical parameters is that until recently the sample of
objects with good optical/UV/X-ray monitoring, having the required cadence of observations to probe the necessary timescales
(up to years) was small. For example, the studies of \citet{wold07} and \citet{kelly09} were based on samples of only $\sim 100$
AGNs. Larger variability studies, on the other hand, have been based on spectroscopically identified AGNs on the Stripe-82
\citep{wilhite08, macleod10, ai10, schmidt10, butler11}, which have $\sim 60$ photometric observations over a decade or on ensemble
AGN studies with few observations \citep{vandenberk04}. Since early AGN monitoring was mainly biased towards bright, color selected, or highly variable
sources the calibration of the aforementioned relations is usually not representative of the whole AGN population. It would also
be highly desirable to compare black hole mass estimations obtained using different methods to assess systematic uncertainties
(e.g., traditional methods vs. optical/X-ray variability black hole mass estimations). We expect to address most of these issues
in the future using data collected as part of our survey.

Since most of the AGN bolometric luminosity is emitted in the UV/optical part of the spectrum and their spectral energy distribution
(SED), this is the region of choice to carry out an AGN survey. However, the UV/optical region
can be subject to strong obscuration. On the other hand, the X-rays --particularly hard X-rays-- can pass through the obscuring material,
thus deep X-ray surveys provide a better census of the AGN population, particularly at low luminosities. X-ray observations are expensive
and therefore surveys at these energies are usually shallow or cover small areas. Clearly, to increase the synergy of any AGN variability
survey it is necessary to carry out intense optical monitoring in fields with extensive X-rays observations.
This point is addressed in our survey by means of monitoring fields with extensive multi-wavelength coverage, ranging
from the X-rays to the mid-IR, and spectroscopic follow-up.

Traditionally the optical selection of AGNs has made use of the fact that AGNs show an UV excess in their SED compared
to stars \citep[see][]{schmidt83}. The UV-excess technique and the more recent selection methods based on optical colors
\citep[see][]{richards02, richards09} are very efficient in finding AGNs in the regions of color space where the AGN
density is higher then the density of stars or galaxies. However, these selection methods based on optical colors miss
a significant fraction of AGNs with peculiar colors (e.g., red QSOs) or QSOs located at a redshift range ($2.5 \leq z \leq 3.0$)
where their optical colors are similar to those of stars \citep[see][]{fan99, richards02, richards09}. On the other hand,
the fact that AGNs are highly variable makes their selection by means of variability a very promising technique to find them
regardless of their colors. This selection method has been successfully used to identify a large number of new QSO candidates
\citep{kozlowski10, schmidt10, macleod11, butler11, palanque11, kim12, graham14}. \citet{butler11} and \citet{palanque11} have shown
that selecting QSOs by their variability can increase considerably the number of QSO candidates in the redshift
range where the colors of stars is similar to those of AGNs.

Since 2010 we have been carrying out an AGN variability survey using the wide-field QUEST camera located on the 1--m ESO-Schmidt telescope
at La Silla Observatory.  The telescope was fully robotized by Yale University, and the QUEST camera was moved from the 48'' Palomar-Schimdt telescope 
to La Silla in 2009. The aims of our survey are: 1) to test and improve variability selection methods of AGNs, and find 
AGN populations missed by other optical selection techniques \citep[see ][]{schmidt10, butler11, palanque11}; 2) to obtain a large number of well-sampled light curves,
covering timescales ranging from days to years; 3) to study the link between the variability properties (e.g., characteristic timescales and amplitude of variation
and other parametric variability characterizations) with physical parameters of the system (e.g., black-hole mass, luminosity, and Eddington ratio). 

In this paper we present the technical description of our survey, and we study the relation of variability with multi-wavelength properties of X-ray
selected AGNs in the COSMOS field. In Section \ref{obs}, we summarize the characteristics of the ESO-Schmidt telescope and the QUEST camera, 
and describe the operation of the robotic telescope. In Section \ref{fields}, we present an overview of our observation fields, 
and we discuss the reasons which make them special for carrying out an AGN variability survey. In Section \ref{reduction},
we present the data reduction steps. In Section \ref{astrometry}, we describe the steps to obtain the astrometric solution for the QUEST
frames, and we present the quality of our astrometry. In Section \ref{photometry}, we define the $Q$-band photometric system, and describe
in detail the steps to obtain well-calibrated light curves, including a nonlinearity correction to the photometry. In Section \ref{photometry},
we also demonstrate the quality of our photometry, compare the quality of aperture and point spread function (PSF) photometry, and present
examples of our light curves. In Section \ref{op_var_vs_multiwave} we study the relation of optical variability with multi-wavelength 
properties of a sample of XMM-COSMOS X-ray selected AGNs. In Section \ref{summary} we summarize our work and present our conclusions.
A standard $\Lambda$ cold dark matter cosmology with $H_{0} = 70$ km~s$^{-1}$~Mpc$^{-1}$, $\Omega_{M}=0.27$, and $\Omega_{\Lambda}=0.73$
is assumed throughout the paper.

\section{TELESCOPE AND CAMERA DESCRIPTION}
\label{obs}

Schmidt telescopes are the instrument of choice for large area sky surveys because of their large field of view. 
The QUEST Camera was designed to operate at the $48"$ Samuel Oschin Schmidt Telescope at the Palomar Observatory \citep[see][]{baltay07}.
Since 2009 the camera has been located in the $1$--m Schmidt telescope of the European Southern Observatory (ESO) at La Silla, Chile 
(see Table \ref{Tel_tab}). This telescope is one of the largest Schmidt configurations in the southern hemisphere, situated in a dry site 
with dark skies and good seeing. Having a nearly identical optical configuration to the Palomar Schmidt, the QUEST camera was installed 
at La Silla without any changes to its front-end optics \citep{rabinowitz12}. The survey uses the Q-band filter described below 
(Section \ref{Q_band_sec}). 

Telescope pointing and camera exposure are coordinated by a master scheduling program \citep{rabinowitz12}. A remote operator of 
one of the larger telescopes at the site (the ESO $3.6$m) decides when conditions are appropriate for opening the telescope, and sends a remote 
command each night to enable the control software to open the dome. The control software automatically closes the dome whenever
another nearby telescope (the $2.2$m) is closed, when the sun rises, or when the remote operator sends a command to close.
The remote operator can monitor and control the state of the Schmidt telescope via a web-based interface.

\begin{figure}
\plotone{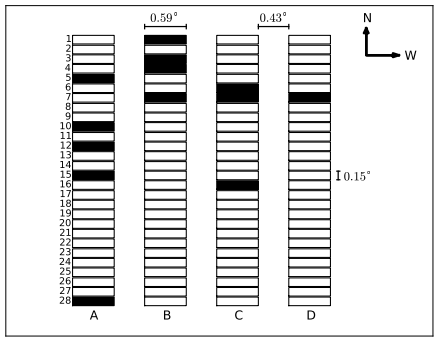}
\caption{QUEST camera array, in black are shown death CCDs. The camera consist in four ``fingers'' of 28 CCDs. The fingers are flagged as
A, B, C, and D, and the columns from 1 to 28. The gaps between adjacent fingers are $0.43^{\circ}$. Each CCD has $2400 \times 600$ pixels
of $0.59^{\circ} \times 0.15^{\circ}$.}
\label{cameralayout_fig}
\end{figure}

The CCD camera is located at the prime focus of the telescope about $3$ m from the primary mirror. Its properties are summarized in 
Table \ref{Cam_tab}. The camera consists of 112 CCDs arranged in four rows or {\it ``fingers''} of 28 CCDs each as shown in Figure 
\ref{cameralayout_fig}, and covers $4.6^{\circ} \times 3.6^{\circ}$ (north-south by east-west) on the sky. The fingers are flagged 
A, B, C, and D and the columns of CCDs from 1 to 28. The gaps between the active areas of the CCDs in a row are typically $1$~mm, 
and the gaps between the active areas in adjacent {\it ``fingers''} are typically $22.8$~mm corresponding to $25\farcm8$. Each
CCD has $600 \times 2400$ pixels of $13~\mu$m~$\times$~$13~\mu$m in size, and a pixel scale of $0\farcs882$ pixel$^{-1}$. Therefore, to obtain a full coverage of the 
$4.6^{\circ} \times 3.6^{\circ}$ field-of-view it is necessary to displace the telescope by half degree in right ascension to obtain an 
image-pair made of two tiles (see Figure \ref{fields_first_fig}).

Several CCDs have areas of high dark current owing to defective, electron-emitting pixels in the CCD substructure or to a bad 
readout amplifier. By subtracting dark calibration images, the unfavorable influence of these defects on source detection is 
largely eliminated. However, about 16\% of the CCDs are useless because they are permanently off, randomly turn on and off, 
or have large defective areas which make it impossible to obtain an acceptable PSF, all of which hamper the astrometric solution 
due to the low number of stars detected and fake detections. The effective sky area covered by the functioning CCDs is $\sim 7.5~ deg^{2}$.

\begin{deluxetable} {ll}
\tablecolumns{2}
\tablenum{1}
\tablewidth{0pc}
\tablecaption{Properties of the ESO $1$--m Schmidt telescope.\label{Tel_tab}}

\startdata
\hline
\hline

Aperture Diameter        & $1$~m \\
Focal Length             & $3.05$~m \\
$f$--ratio               & $3.05$ \\
Plate scale              & $14.74~\mu$m$/~\arcsec$ \\
Latitude of Observatory  & $-29^{\circ}:15'$ \\
Longitude of Observatory & $70^{\circ}:44'$ \\
Elevation                & $2,375$~m \\

\enddata
\end{deluxetable}

\begin{deluxetable} {ll}
\tablecolumns{2}
\tablenum{2}
\tablewidth{0pc}
\tablecaption{Properties of the QUEST Camera.\label{Cam_tab}}

\startdata
\hline
\hline

Number of CCDs           & $112$ \\
CCD Pixel Size           & $13~\mu$m $\times~13~\mu$m \\
Number of Pixels per CCD & $600 \times 2,400$ \\
Pixel Size on Sky        & $0\farcs882 \times 0\farcs882$ \\
Array Size, CCDs         & $4 \times 28$ \\
Array Size, Pixels       & $9,600 \times 16,800$ \\
Array Size, cm           & $19.3$ cm $\times$ $25.0$ cm \\
Array Size on Sky        & $3.6^{\circ} \times 4.6^{\circ}$ \\
Sensitive Area           & $9.6$ square degrees \\
Total Pixels             & $161 \times 10^{6}$ \\

\enddata
\end{deluxetable}

\section{SURVEY FIELDS}
\label{fields}

We carried out the AGN variability survey observing the COSMOS, ECDF--S, ELAIS--S1, XMM--LSS and Stripe-82 fields. 
These fields were chosen due to the wealth of ancillary data available for them, including {\it XMM--Newton}, {\it Chandra},
{\it GALEX}, {\it HST}, {\it Herschel}, {\it Spitzer}, and ground-based photometry and spectroscopy.  

Additionally, the COSMOS, ECDF--S, ELAIS-S1, and XMM--LSS fields have been repeatedly observed in the near-IR since 2009, as part of the 
VISTA public surveys UltraVISTA \citep{mccracken12} and VIDEO \citep{jarvis13}. The aims of these VISTA surveys are to study the evolution 
of galaxies out to $z\sim4$, and achieve a comprehensive view of AGNs and the most massive galaxies up to the epoch of re-ionization.
UltraVISTA observations were carried out over the same period as our survey, and an article presenting the study of the AGN near-IR
light curves is underway \citep{sanchez15}.

The equatorial Stripe-82 is particularly valuable for variability studies \citep[see][]{sesar07}. This region 
was repeatedly observed as part of the Sloan Digital Sky Survey SDSS since 1998, this will allow us to extend
light curves of some of Stripe-82 AGNs combining  SDSS and QUEST photometry to more than 15 years. 

\begin{figure*}
\plottwo{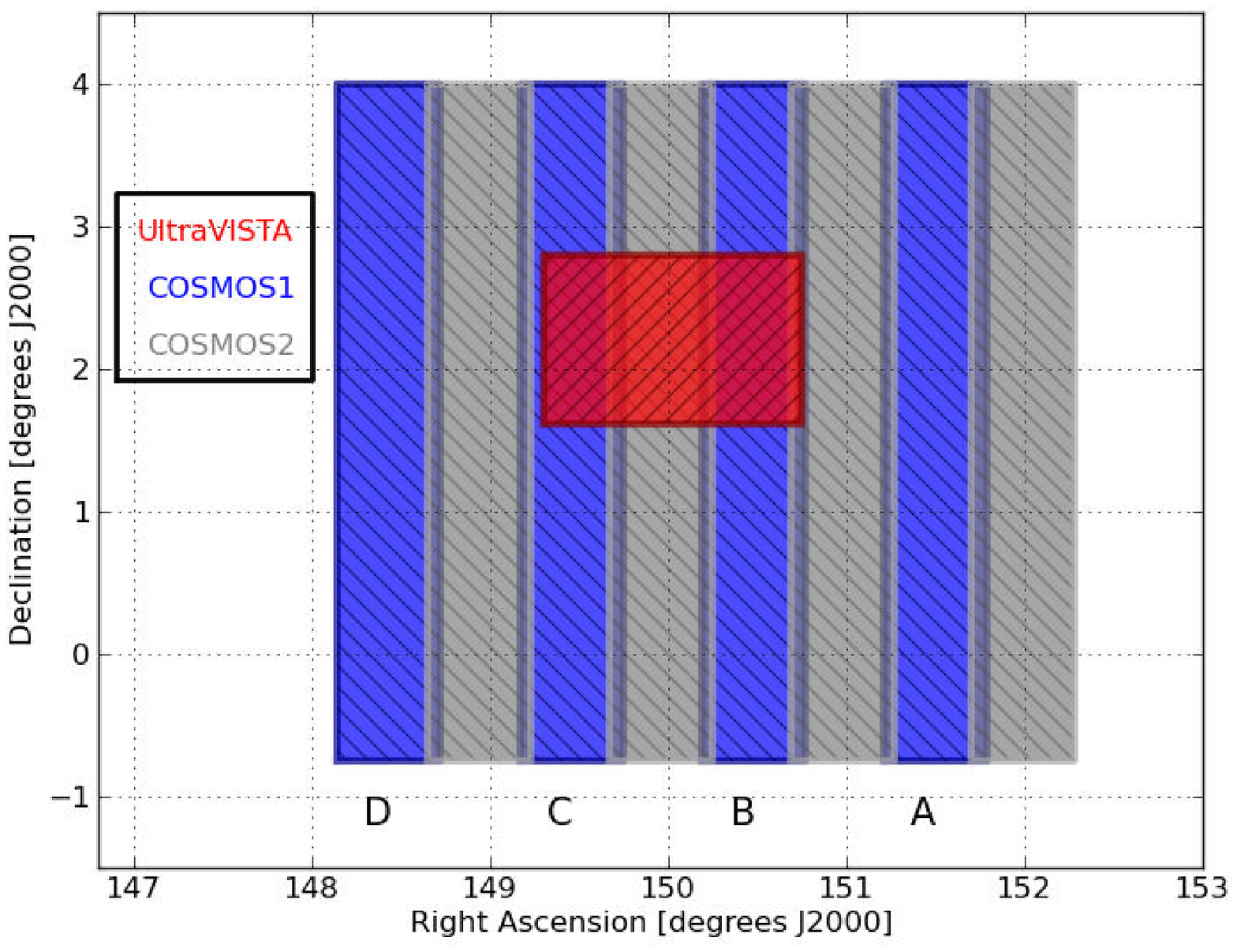}{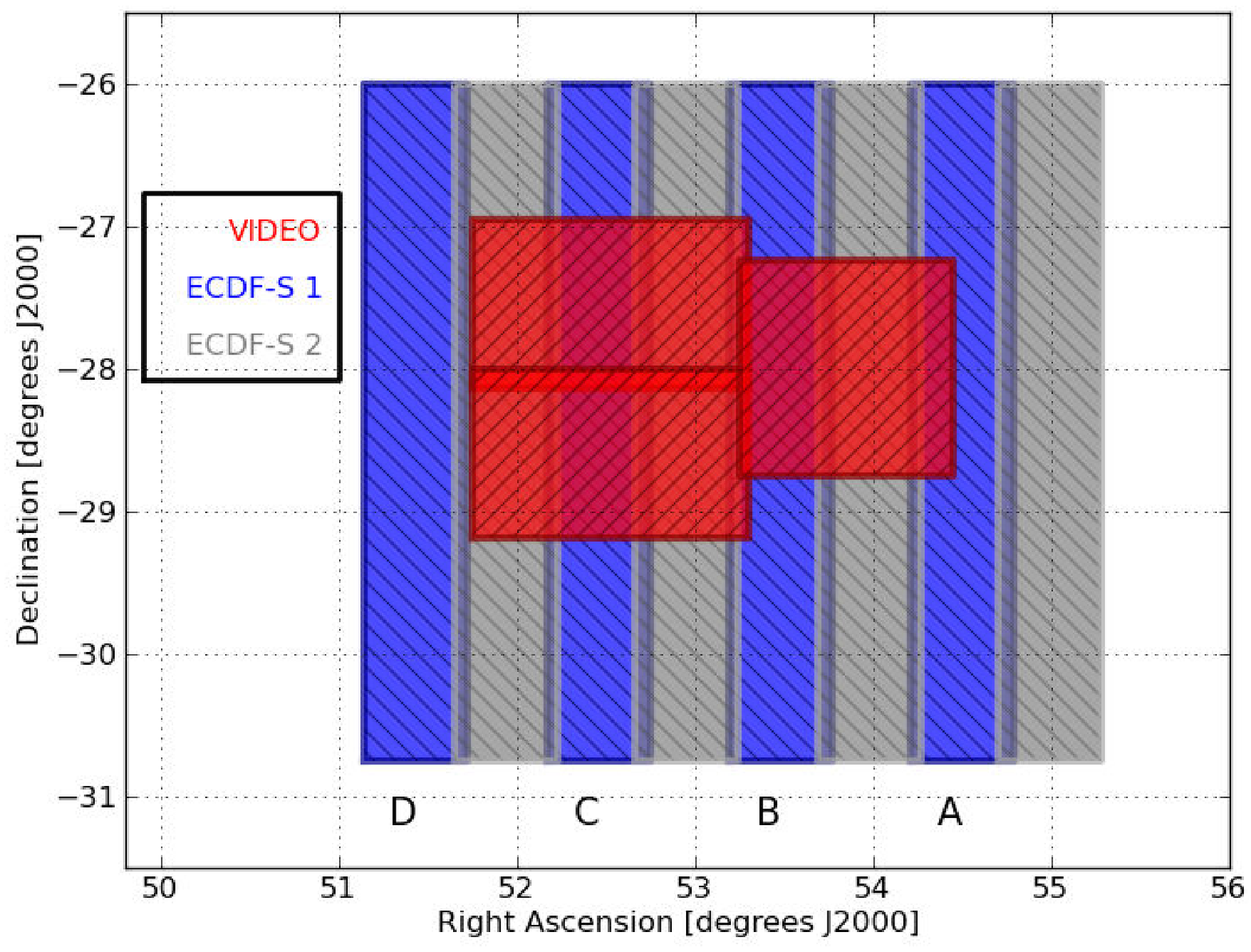}
\caption{Observing layout for the COSMOS (left) and ECDF-S (right) fields. In blue it is shown COSMOS1 and ECDF-S 1 (first tile) and in grey COSMOS2 and
ECDF-S 2 (second tile). The second tiles are displaced by half degree from the first tiles. In red we show the area repeatedly observed by ultraVISTA
and VIDEO in the COSMOS and ECDF-S fields, respectively. We flag the position of the four (A, B, C, and D) fingers for the first tiles.}
\label{fields_first_fig}
\end{figure*}

\begin{figure*}
\plottwo{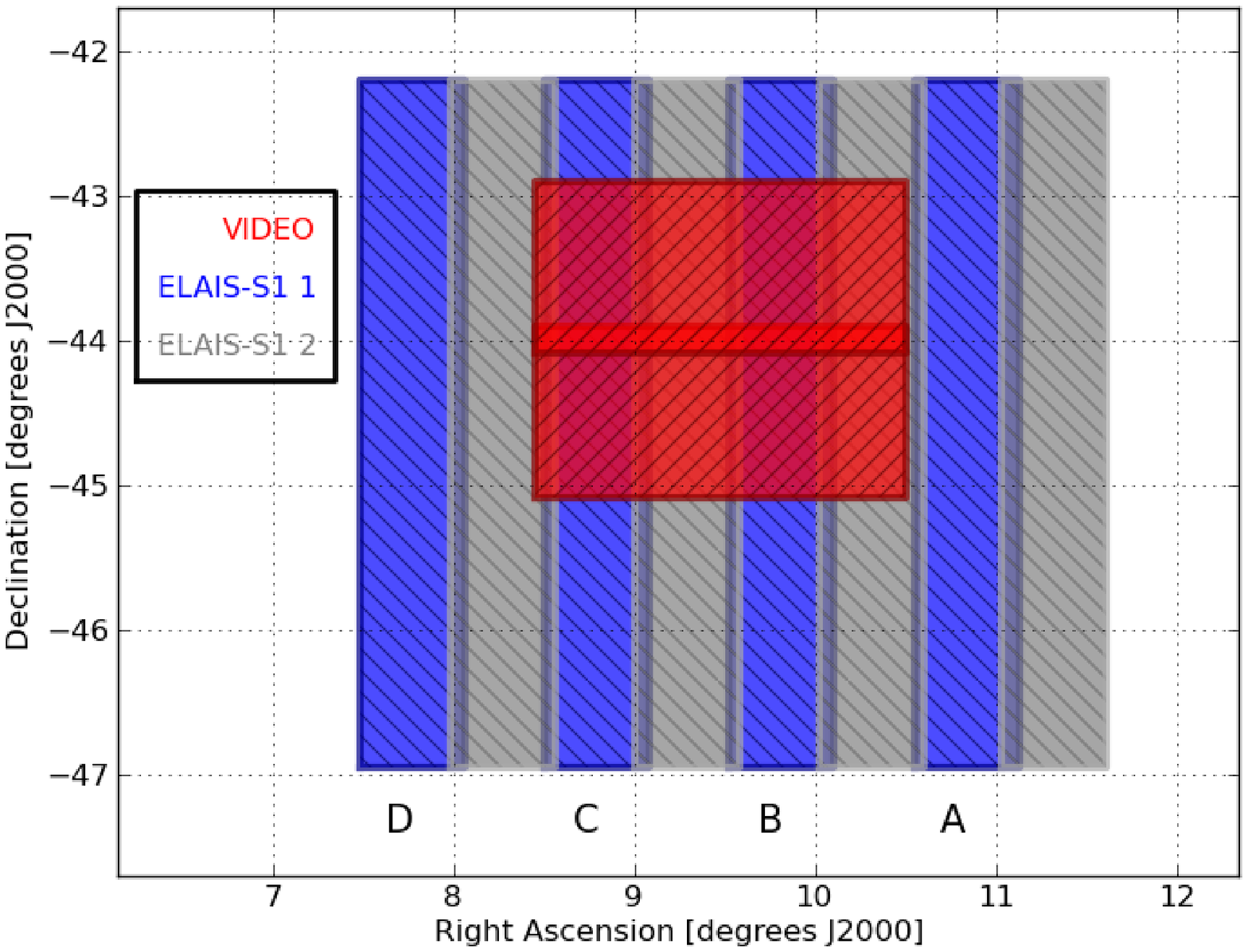}{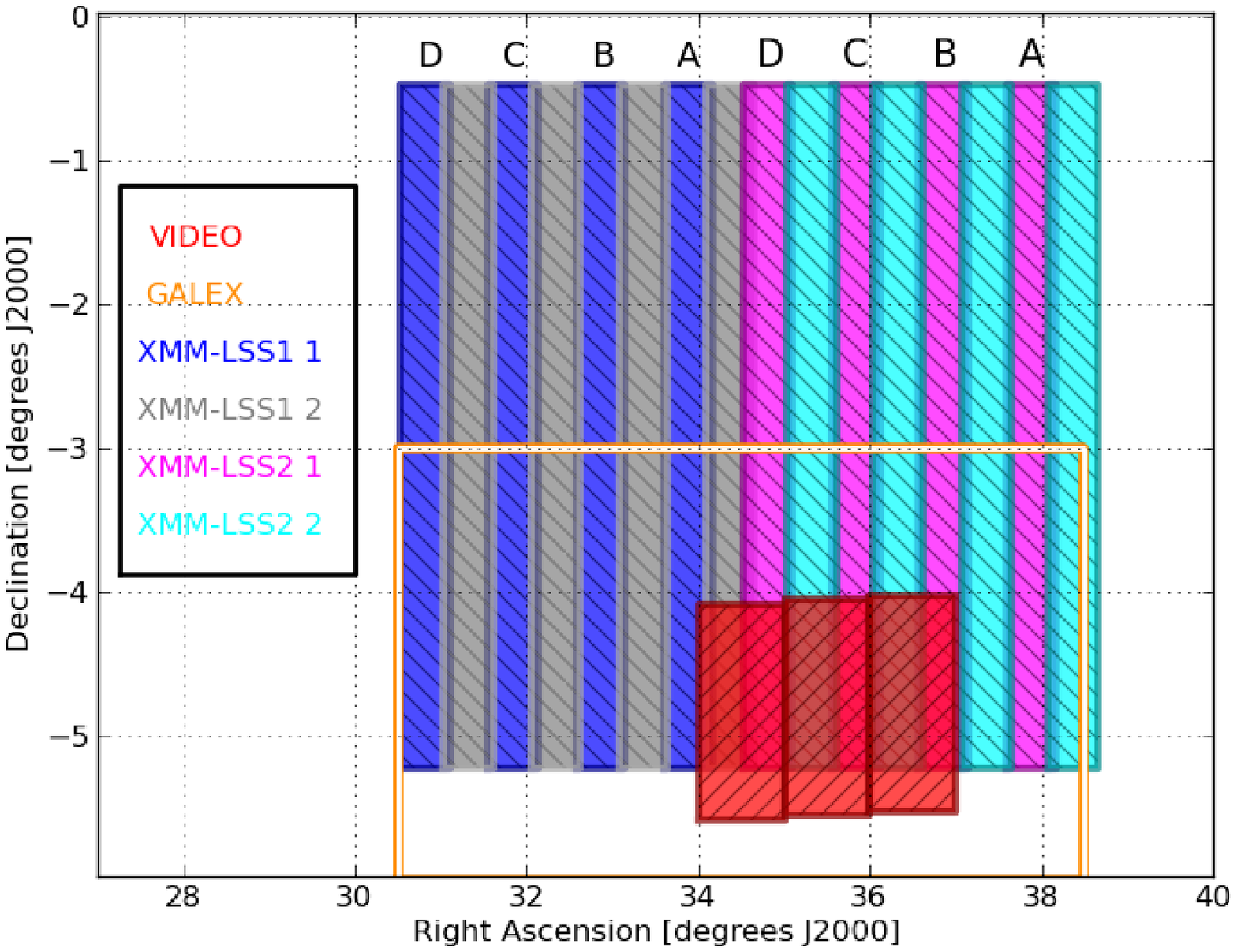}
\caption{Observing layout for the ELAIS-S1 (right) and the XMM--LSS (left) fields. In blue are shown ELAIS-S1 1 and XMM--LSS1 1 (first tile),
in grey ELAIS-S1 2 and XMM--LSS1 2 (second tile), in magenta XMM--LSS2 1 (third tile), and in cyan XMM--LSS2 2 (fourth tile). Each tile is
displaced by half degree from the previous. In red we show the area repeteadly observed by VIDEO, and we flag the position of the four (A, B, C, and D)
finguers for the first tile.}
\label{fields_second_fig}
\end{figure*}

In Figures \ref{fields_first_fig} and \ref{fields_second_fig} we show the layout used to cover the COSMOS, ECDF-S, ELAIS-S1
and XMM--LSS fields. For the COSMOS, ECDF-S, ELAIS-S1 and Stripe-82 we used two tiles to cover the entire regions,
while to cover XMM--LSS field we used four tiles (see Figure \ref{fields_second_fig}).

A brief summary of the observations is as follows. In 2010 we obtained just a few observations for some fields
($\sim 20$ nights). During 2011--2012, due to problems with the dome wheels of the telescope, we observed between
$100$ to $150$ nights per field (twice per night). From 2013 onwards the observations have been performed more regularly using
an exposure time of $180$ seconds (twice per night) reaching a magnitude limit of $Q \sim 20.5$ ($r \sim 21$), and observing each
of our fields in more than 100 nights per year. Between March of 2011 and the end of December 2014 we observed COSMOS 
on $\sim 370$ nights, ECDF-S on $\sim 290$ nights, ELAIS-S1 on $\sim 500$ nights, Stripe-82 on $\sim 450$ nights
and XMM--LSS on $\sim 360$ nights. In the case of XMM--LSS we usually observed 2 tiles per night, either tile 1 and
tile 2 or tile 3 and tile 4. Observations of ECDF-S and XMM-LSS until 2012 have been reduced, while ELAIS-S1 and Stripe-82 
are reduced until 2014. We expect to continue collecting data until mid 2016. 

From Section \ref{astrometry} onwards, the analysis presented corresponds to data obtained from 2010 to 2012 in the COSMOS field.

\section{DATA REDUCTION}
\label{reduction}

For a typical night we take darks of 10, 60, and 180 seconds, as well as morning and evening twilight flats. The exposure times
for our science images are either 60 or 180 seconds, with two image-pairs obtained per field per night. 
We reduce the QUEST data using our own custom IRAF\footnote[4]{IRAF is distributed by the National Optical Astronomy Observatories, which are 
operated by the Association of Universities for Research in Astronomy, Inc., under cooperative agreement with the National Science Foundation.} 
scripts that carry out the reduction steps described below. 

Darks are combined to obtain a master dark for each detector and each exposure time. Then we subtract the master
dark of the appropriate exposure time from the science images. The pixel-to-pixel variations of the science 
frames are corrected dividing them by the master flats obtained as we describe below. Finally, bad pixels in the science frames 
are interpolated using the IRAF task FIXPIX, which requires a bad pixel mask as input. We constructed bad pixel masks for each
detector. To construct the bad pixel mask we use the IRAF task CCDMASK, taking as input the ratio between a high count twilight flat 
and a low count twilight flat.

Given the large size of the camera and the large field of view of the telescope, to obtain dome flats using a uniformly illuminated panel 
was unfeasible. Furthermore, the automatic operation of the telescope complicates a careful acquisition of dome flats. Instead twilight 
flats were preferred because they can be easily acquired automatically. Typically twilight flats are taken at the beginning (evening flats), 
and the end (morning flats) of the observing night. On many occasions flats were not useful because of moon illumination gradients, a closed dome, 
or cloudy conditions.

To avoid these problems and to obtain a good pixel-to-pixel variation correction we selected, from dark subtracted and trimmed twilight flats, 
the flats with mean counts above the average obtained over approximately two weeks of observations. Then we median combine these twilight flats and 
normalize them to obtain master flats for each detector. Typically we combined more than 50 twilight flats per detector.

\section{ASTROMETRY}
\label{astrometry}

\subsection{Astrometric Solution}

The frame headers lack basic astrometric information. To solve for this, we have created basic astrometric headers which contain 
a crude approximation to the true World Coordinate System information (WCS) of the frame. We introduce the estimated header keys (i.e., an
estimated WCS) into the image headers using MissFITS\footnote[5]{http://astromatic.net/software/missfits}. Then we proceed to obtain a first good 
astrometric solution using astrometry.net\footnote[6]{http://astrometry.net/} \citep{lang10}. This software provides a fast and robust method 
to calibrate astronomical images, and is able to produce a {\it blind astrometric calibration} of the QUEST frames. The output is an 
image with WCS information which can be further refined. 

The solution produced by astrometry.net is refined using the Software for Calibrating AstroMetry and Photometry \citep[SCAMP,][]{bertin06}.
SCAMP is a program that computes precise astrometric projection parameters from source lists obtained directly from FITS images.
The input lists for SCAMP are in SExtractor binary format \citep[``FITS\_LDAC'',][]{bertin96}, and must contain the centroid 
coordinates, centroid errors, astrometric distortion factors, flux measurements, and flux errors. Furthermore SCAMP creates frame headers 
ready to be used in an image stacking process. Both astrometry.net and SCAMP used the reference catalog USNO--B \citep{monet03}, which has an 
astrometric accuracy of $0\farcs2$.

\subsection{Assessment of QUEST astrometry}

To test the precision of our astrometric solution we assessed the internal astrometry by computing
the standard deviation of the internal cross-matching of sources. To assess the absolute astrometric precision we 
cross-matched our source catalogs with SDSS sources, and then we computed residuals of the cross-matching.

\subsubsection{Internal QUEST Astrometric Consistency}
\label{internal_astrometry}

To study the astrometric consistency we defined two samples of point-like sources: the subsample called standard stars and the subsample of
variable objects. We used a $\chi^{2}_{LC}$ statistics 
\citep[where LC stands for light curve, generated as detailed in Section \ref{lc_generation_sec}; see][]{sesar07} to distinguish between 
standard stars and variable objects, which is defined as follows:

\begin{equation}
\chi^{2}_{LC}~= {1 \over (n-1)} \sum_{i=1}^{n} {(m_{i} - \bar{m})^{2} \over \sigma_{i}^{2}} 
\label{chi_squared_eq}
\end{equation}

\noindent where $n$ is the number of detections, $m_{i}$ is the magnitude, $\bar{m}$ is the mean magnitude, and $\sigma_{i}$ is
the photometric error. The standard stars subsample consists of objects with $n$~\textgreater~$10$, a standard deviation in 
its light curve ($\sigma_{LC}$) smaller than the median error observed in the photometry ($\sigma_{median}$), $\sigma_{LC}$ \textless~$0.035$ mag,
and $\chi^{2}_{LC}$ \textless~$1.5$. The other group called variables contains objects with $n$~\textgreater~$5$, 
$\sigma_{LC}$~\textgreater~$2 \times \sigma_{median}$, $\sigma_{LC}$~\textgreater~$0.05$ mag, and $\chi^{2}_{LC}$~\textgreater~$2.0$. 
As an example we obtained roughly $19,800$ standard stars in our COSMOS field, of which $18,265$ have SDSS counterparts with good photometry. 
Of the $2,790$ variables objects in our COSMOS field roughly $2,500$ have a SDSS counterparts with good photometry. 

We computed the mean and the median dispersion in the Equatorial (J2000) position of the standard stars and obtained $0\farcs08$ and $0\farcs07$, 
respectively. The standard deviation for this subsample was $0\farcs04$. For the variable sub-sample the mean and the median were 
$0\farcs15$ and $0\farcs11$, respectively. The standard deviation for the variables was $0\farcs10$. The variable subsample has many 
bright variable stars, and relatively faint transients. When the bright variable stars brighten, the PSF of these bright stars 
is close to saturation, such that the centroid becomes more uncertain. Something similar happens with faint transients where the centroid 
is more uncertain due to background fluctuations. In addition, after inspection of the light curves we found a significant number 
of non-variable sources classified as variables due to matching with nearby artifacts, thus leading to an increase in the 
astrometric dispersion. Overall our internal astrometric precision is typically $\sim 0\farcs1$.

\begin{figure}
\plotone{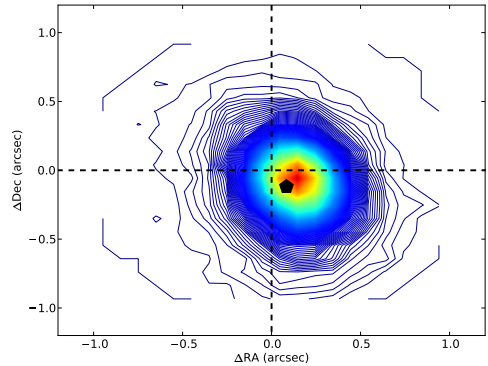}
\caption{Astrometric comparison between QUEST and SDSS positions for all point sources in our COSMOS field. The figure shows the
contour level lines of the number objects as a function of $\Delta$RA and $\Delta$Dec, where $\Delta$RA corresponds to
$\alpha_{SDSS} - \alpha_{QUEST}$, and $\Delta$Dec to $\delta_{SDSS} - \delta_{QUEST}$. The black dashed lines show the
zero offset axis, and the black pentagon marks the position of mean offset.}
\label{quest_vs_sdss_astmetry_all_fig}
\end{figure}

\subsubsection{QUEST vs. SDSS Astrometric Consistency}

To assess our overall accuracy we cross-matched our QUEST subsample catalogs with SDSS stars. For the standard subsample we obtained a 
mean of $0\farcs21$ with a standard deviation of $0\farcs13$, and the median was $0\farcs18$. In the case of the variable group we obtained a 
mean of $0\farcs23$ with a standard deviation of $0\farcs17$, and a median of $0\farcs19$. As before some of the variable sources may be 
contaminated by matching with artifacts thus increasing the dispersion in their positions.

\begin{figure}
\plotone{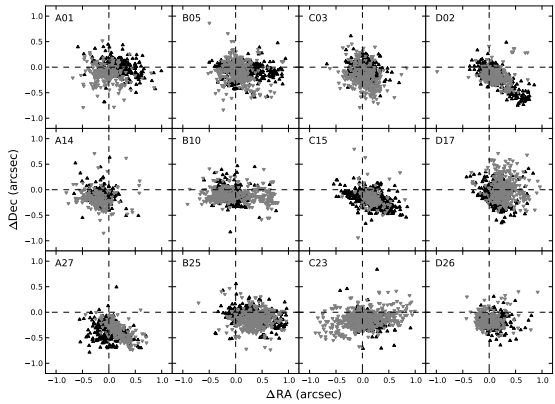}
\caption{Astrometric comparion between QUEST and SDSS positions for twelve randomly selected detectors. In black we show tile one
and in grey tile two. The black dashed lines show the zero offset axis. Overall the mean offsets are within $0.2''$ which is consistent
with the astrometric accuracy of USNO-B \citep{monet03}.}
\label{quest_vs_sdss_astmetry_exam_fig}
\end{figure}

In Figure \ref{quest_vs_sdss_astmetry_all_fig} we show an astrometric comparison between QUEST positions and SDSS positions
for all point sources in our COSMOS field. We found mean offsets of $\alpha_{SDSS} - \alpha_{QUEST}=0\farcs08$ with a dispersion of $0\farcs25$,
and $\delta_{SDSS} - \delta_{QUEST}=-0\farcs12$ with a dispersion of $0\farcs19$. We found that these offsets are variable across the field
(see Figure \ref{quest_vs_sdss_astmetry_exam_fig}), and possibly are related to small systematic differences between SDSS and USNO--B,
since both astrometry.net and SCAMP used as reference catalog USNO--B \citep{monet03} which has an accuracy of $0\farcs2$. Therefore, 
the differences in the positions between QUEST and SDSS objects are within the expected astrometric accuracy of the reference
catalogs.

\section{PHOTOMETRY}
\label{photometry}

\subsection{$Q$-band Photometric System}
\label{Q_band_sec}

\begin{figure}
\plotone{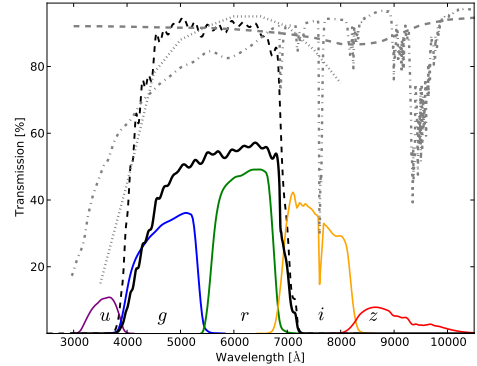}
\caption{System responses of the SDSS bandpasses (in colors) and the Q--band (continuous black) at an airmass of $1.3$. We
computed the $Q$-band system response profile from the multiplication of the QUEST filter \citep[dashed black, see][]{rabinowitz12},
twice the mirror reflectivity (dashed grey), the sky transmission at an airmass of $1.3$ (dotted-dashed grey), and the camera
quantum efficiency \citep[dotted grey, see][]{baltay07}. We assume a flat throughput of $90$\% for the optics of the camera.
The Q--band filter covers a spectral region similar to a $(g+r)_{SDSS}$ filter.}
\label{qfilter_fig}
\end{figure}

The QUEST-La Silla survey uses a broad filter covering from $4000$~\AA~to $7000$~\AA~called the $Q$-band. This bandpass was designed to avoid 
the fringing often present in the images taken as part of the Palomar--QUEST survey \citep{baltay07}. In Figure \ref{qfilter_fig} we show an 
estimation of the system effective response of the $Q$-band (continuous black line). We computed the $Q$-band system response profile 
using the QUEST filter \citep[dashed black line, see][]{rabinowitz12}, the mirror reflectivity (dashed grey line), 
the Cerro Tololo Inter-American Observatory (CTIO) sky transmission at an airmass of 1.3 (dot-dashed grey line), and the 
quantum efficiency of the camera \citep[dotted grey line, see][]{baltay07}. We assume a flat throughput of $90$\% for the optics 
of the camera. For comparison we show the filter response curves of SDSS\footnote[7]{http://classic.sdss.org/dr7/instruments/imager/filters/}
at an airmass of 1.3. It can be observed that the $Q$-band system response used in La Silla-QUEST is similar
to a broad $(g+r)_{SDSS}$ filter. Therefore, we decided to calibrate the $Q$-band performing differential photometry using reference
stars from SDSS-DR7 \citep{abazajian09} in the COSMOS field.

To calibrate our photometry we created a catalog of $Q$ magnitudes of SDSS point sources with good $g$ and $r$ photometry.
To make this catalog we transformed $g$ and $r$ magnitudes to flux, and then we added these fluxes to obtain an equivalent to the
flux in $Q$. This flux was transformed to $Q$ magnitudes in the AB system.

\subsection{PSF Photometry}

We carried out PSF photometry in the QUEST frames using custom scripts that run DAOPHOT \citep{stetson87} 
on each epoch. We decided to compute PSF photometry, since for faint stellar sources and crowded fields PSF 
photometry is usually better than aperture photometry. However, PSF photometry is more expensive in computing time. Besides,
we find that occasionally QUEST frames can yield a bad PSF due to telescope tracking problems, bad weather, moonlight 
scattered in to the frames, or frames with too many cosmetic problems.

\subsection{Aperture Photometry}

We also carried out aperture photometry using SExtractor \citep{bertin96}. SExtractor is a software to compute aperture 
photometry, and other source parameters (e.g., stellarity, source shape, and photometry quality) quickly.

To obtain AGN aperture photometry with minimal contribution from the host galaxy we determined an optimal 
aperture of $\sim 7$ pixels ($\sim 6\farcs18$). This aperture is $2.5$ to $3.0$ times wider than the typical
seeing of the QUEST--La Silla images, which ranges between $2\farcs0$ and $2\farcs5$, and is relatively 
insensible to typical seeing variations at La-Silla observatory site.

\subsection{Comparison between Aperture Photometry and PSF Photometry}

\begin{figure*}
\plottwo{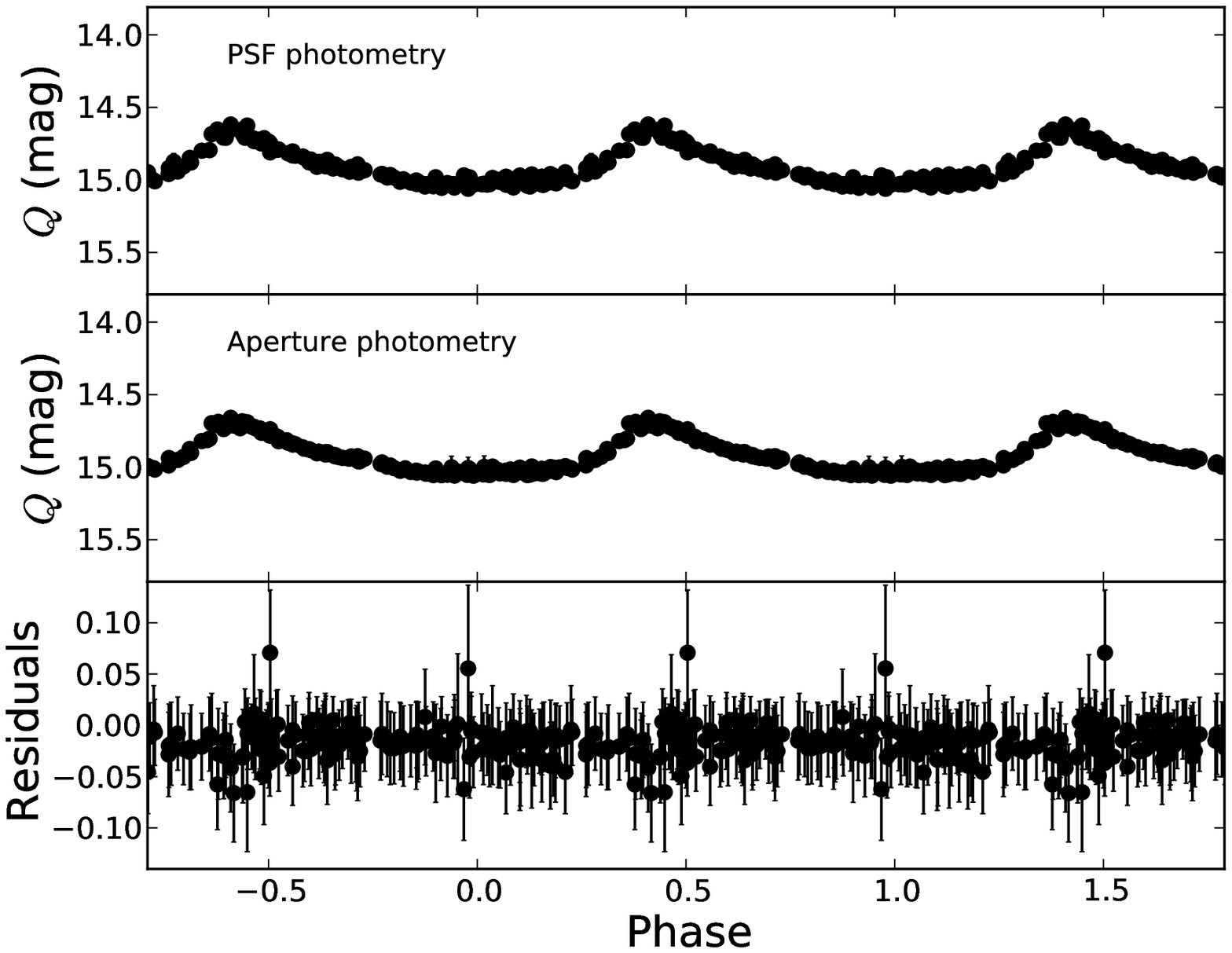}{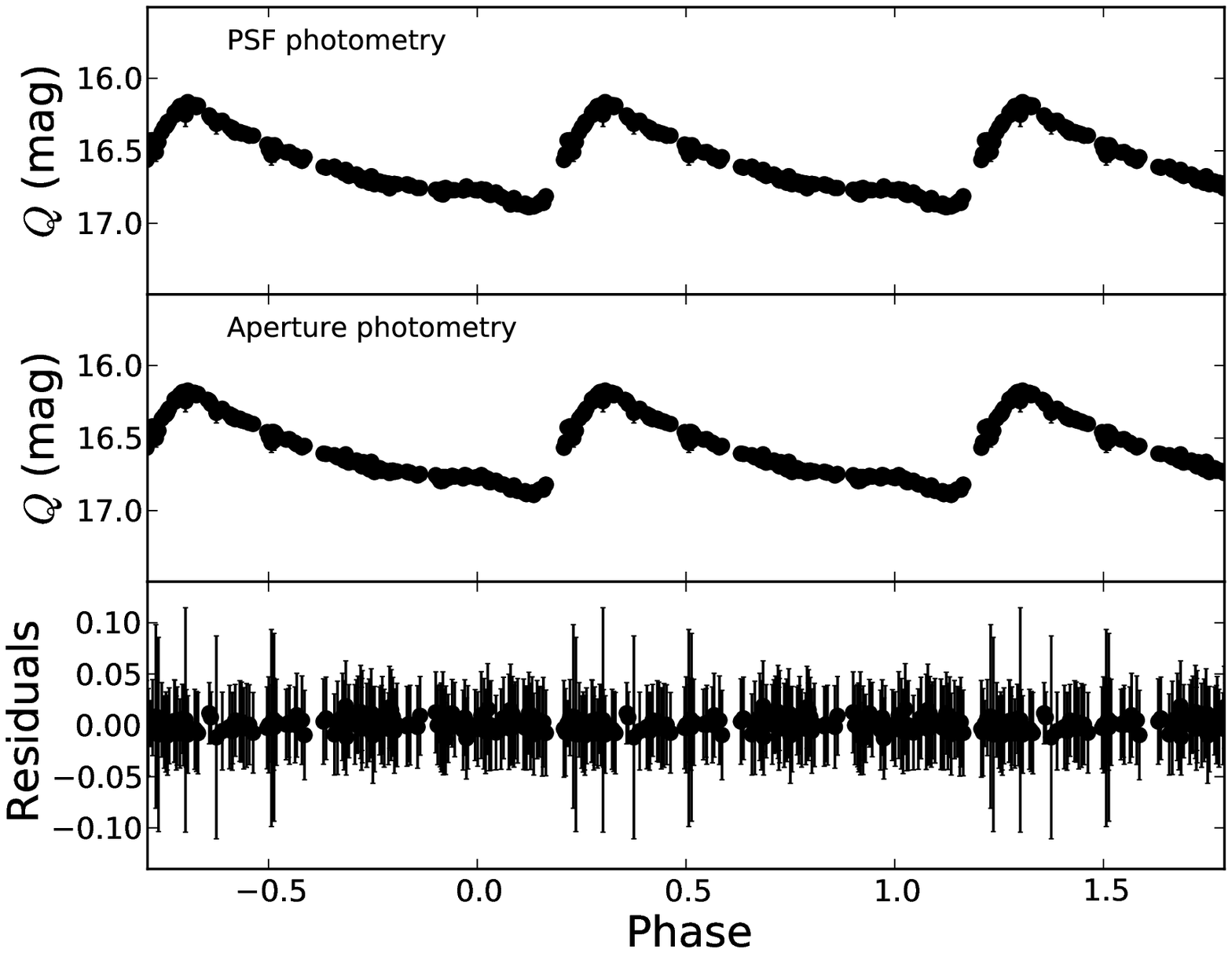}
\caption{Examples of periodic variable stars in the QUEST-COSMOS field, the left panel star has a period of $\sim 1.44$ hrs,
while the right panel object has a period of $\sim 14.59$ hrs. We present PSF and aperture photometry in the top and middle
panels, respectively. In the bottom panel we show the residuals of the difference between PSF and aperture photometry.}
\label{var_exam_fig}
\end{figure*}

Overall we found that PSF and aperture photometry agree very well and the differences are within the uncertainties.
The mean difference between aperture and PSF photometry is $0.02$ mag, with a standard deviation of $0.03$ mag. In
Figure \ref{var_exam_fig} we compare PSF and aperture photometry for variable stars in the QUEST-COSMOS field.

After a random inspection of the light curves (see Section \ref{lc_generation_sec}) we found that PSF photometry achieves
slightly better results for faint point sources, whilst for the rest of the objects both techniques yield similar results.
However, currently our pipeline fails to obtain a PSF solution for a significant number of image frames. Hence, for the
AGN variability analysis of Section \ref{op_var_vs_multiwave} we use aperture photometry. In the future we might be able
to improve our PSF photometry using ad hoc parameters for each detector.

\subsection{Linearity Correction}
\label{lincorr_sec}

\begin{figure*}
\plottwo{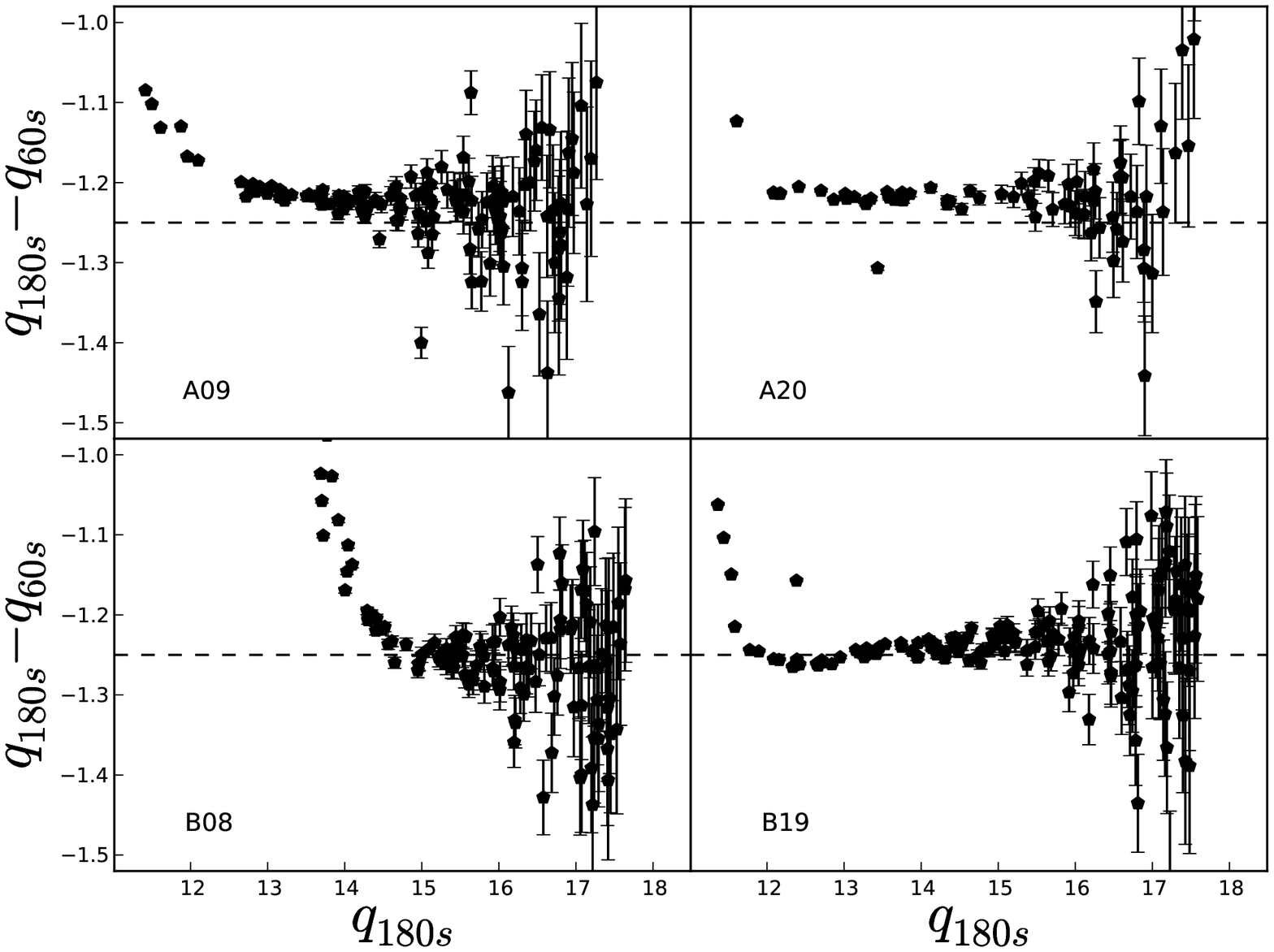}{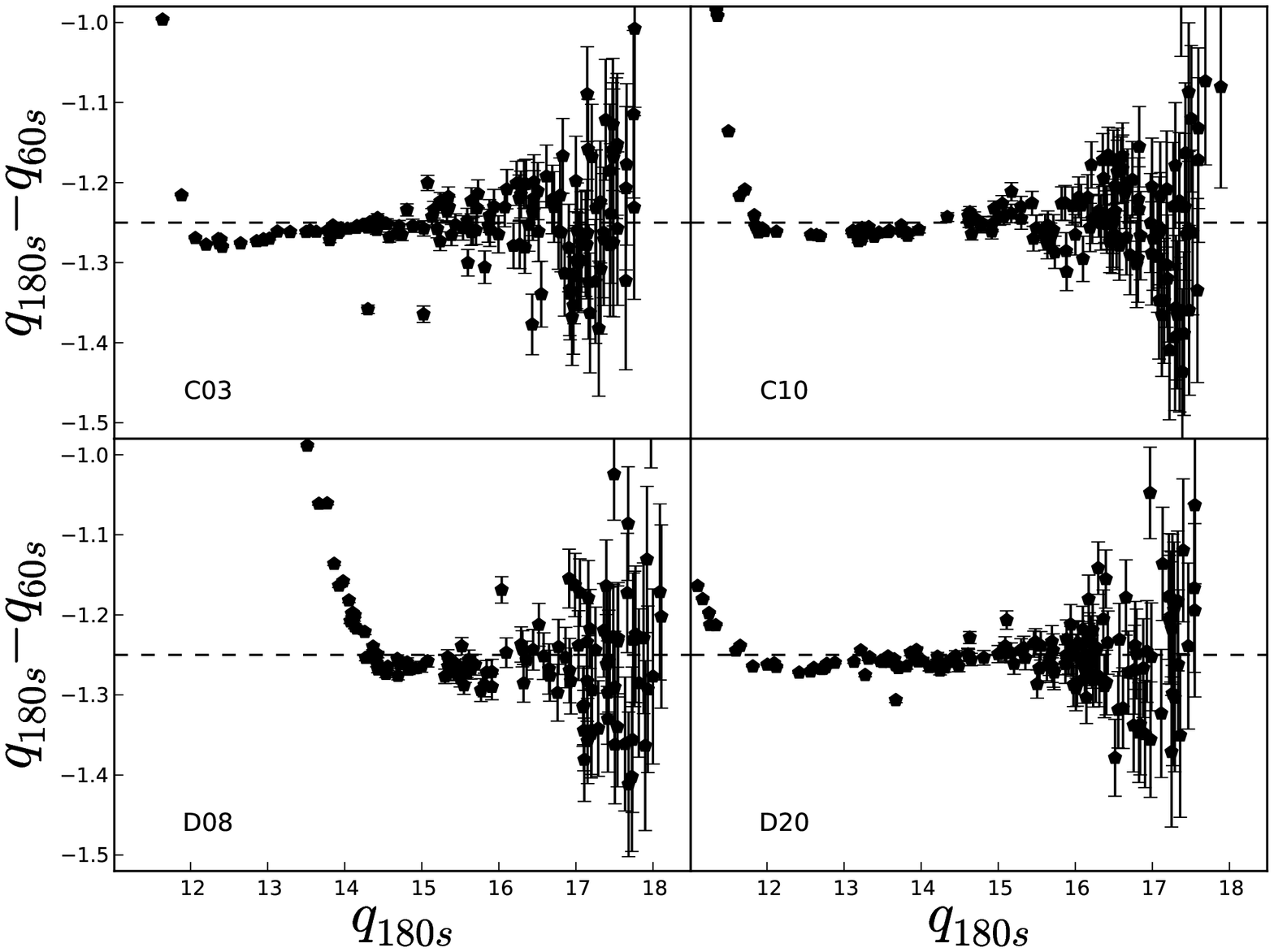}
\caption{Residuals of the difference between $180$~seconds instrumental magnitudes ($q_{180s}$) and $60$~seconds instrumental magnitudes ($q_{60s}$) as
a function of $q_{180s}$ for the detectors of Figure \ref{zp_plot_fig}. The $180$~seconds and the $60$~seconds observations were taken on the
same clear night $\sim 5$ min apart. These examples show how each detector has its own non--linear behavior. We plot black horizontal
dashed lines to guide the eye.}
\label{180s_60svs180s_fig}
\end{figure*}

\begin{figure*}
\plottwo{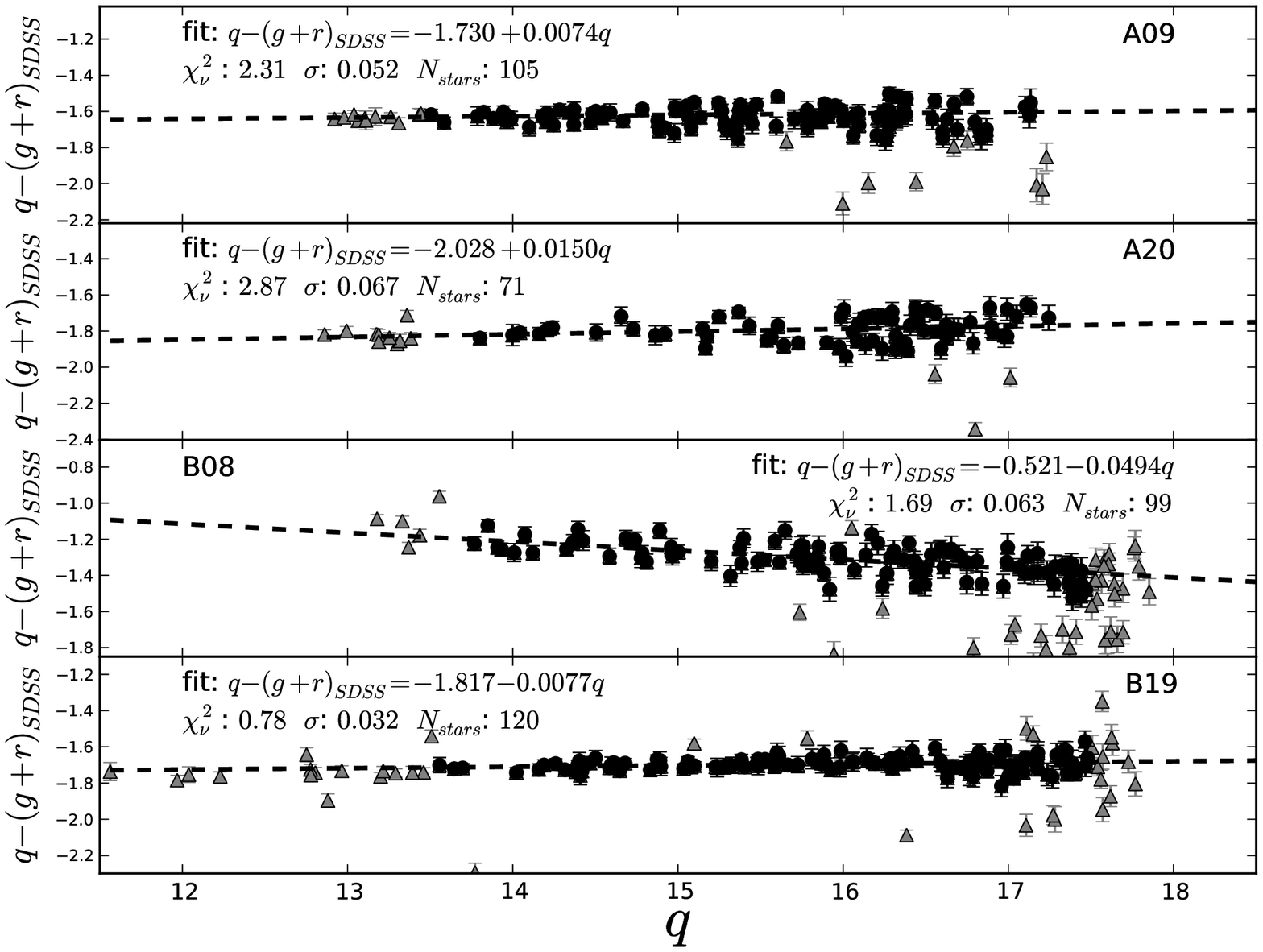}{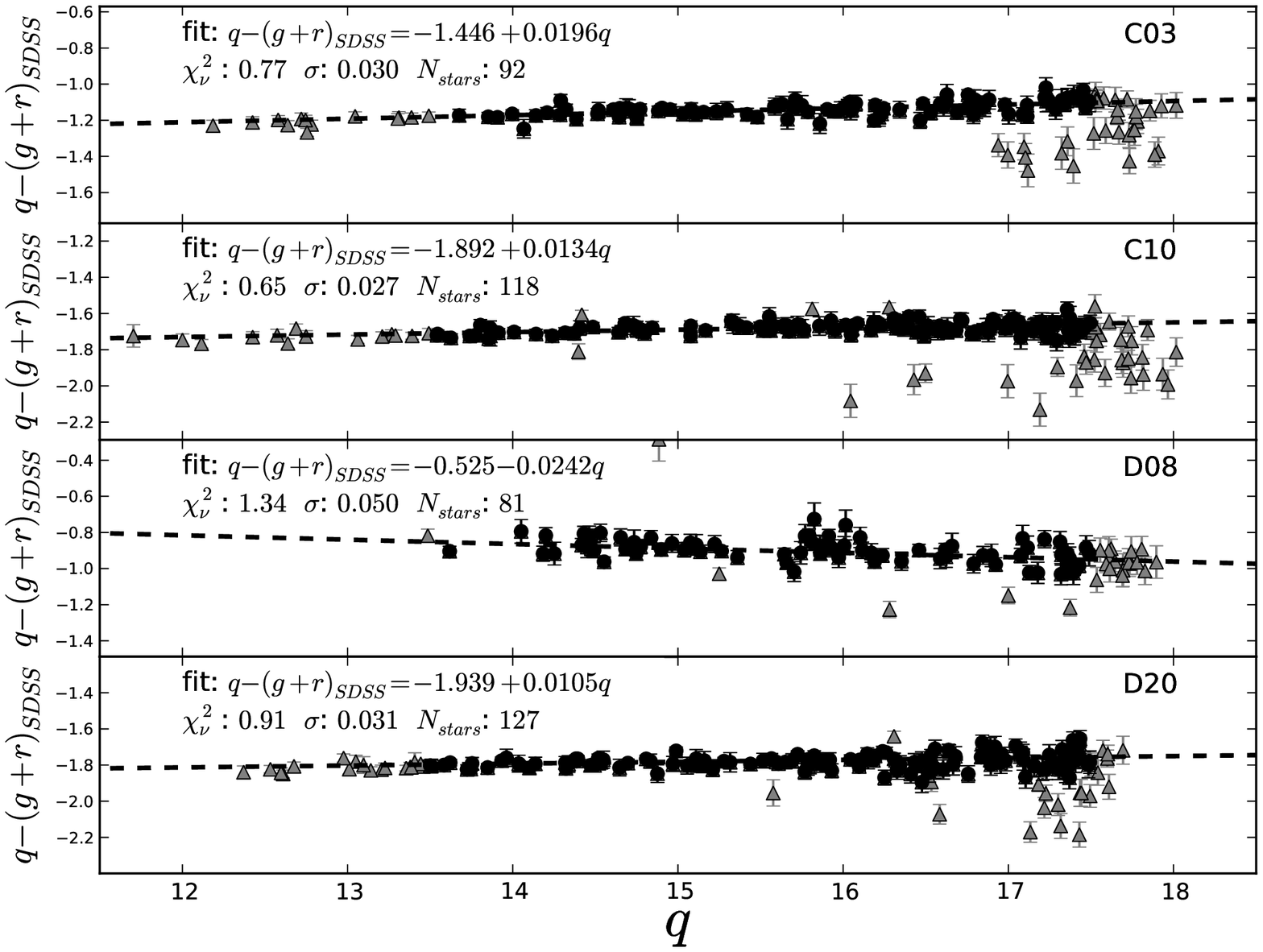}
\caption{Residuals of the difference between the QUEST instrumental magnitudes ($q$) and the $(g+r)_{SDSS}$ magnitudes
for stars plotted as a function of $q$. The black points correspond to the data used to compute the photometric zero-point
as function of $q$ (dashed--black line). Grey triangles are point sources with $q$ \textless $17.5$, or $q$ \textgreater 13.5,
or are outliers to the fit and they were not used to compute the best linear fit. On the top of each panel
we give the best fit parameters and the detector name.}
\label{zp_plot_fig}
\end{figure*}

To ensure that non-linearity is not a dominant effect in the calibration of our data, we looked at observations of 60 seconds and 
180 seconds taken close in time and on clear nights. In the perfect case where non-linearities and saturation were unimportant we 
would expect to see flat residuals for the difference between 180 seconds instrumental magnitudes ($q_{180s}$) and the 60 seconds instrumental 
magnitudes ($q_{60s}$). In Figure \ref{180s_60svs180s_fig} we show the residuals of $q_{180s} - q_{60s}$ as a function of $q_{180s}$.
In the plots all detectors show a clear upward trend toward the left corresponding to bright objects.
After an inspection of the images we found that the vast majority of the objects were not saturated, and the most likely explanation
for this behavior is non-linearity of the detector. 

Furthermore, the onset of the non-linearities depends on each detector. For example, in Figure 
\ref{180s_60svs180s_fig} detector A20 has no significant slope, whilst detector C03 shows a clear positive slope.
A pronounced slope is indicative that non-linearity is present.

To correct the non-linearities we inspected the residuals of the difference between the $q$ instrumental magnitudes and 
the $(g+r)_{SDSS}$ magnitudes as a function of $q$. In Figure \ref{zp_plot_fig} we show some examples of $q - (g+r)_{SDSS}$
as function of the instrumental magnitude. For most of the examples a linear fit as function of $q$ is significantly
better representation of the residuals than a constant zero point. For this reason we decided to use a linear expression of
the following form to fit the residuals: 

\begin{equation}
q~-~(g+r)_{SDSS}~=\alpha~+~\beta \times q.
\label{res_eq}
\end{equation}

\noindent To avoid distortions in the fit due to very bright and faint sources we only used objects in the interval 
$13.5$ \textless $q$ \textless $17.5$ (shown in black), and we sigma-clipped the fit to eliminate outliers (grey triangles).
Figure \ref{zp_plot_fig} shows that this model produces good fits to the residuals, even to the data points that were
not used to compute the linear model. 

We determine the reduced $\chi^{2}$ ($\chi^{2}_{\nu}$), the standard deviation ($\sigma$), and the number of stars used to
compute the fit. Subsequently we use these parameters to clean our photometry from bad epochs when constructing the light
curves. We discard photometry with  $\chi^{2}_{\nu}$\textgreater$10$, $\sigma$\textgreater$0.1$, or photometry that was
obtained with less than 20 stars in the calculation of the zero-point. (see Section \ref{lc_generation_sec}). The final
calibrated $Q$-band magnitude is computed as: 

\begin{equation}
Q=q \times (1 - \beta) - \alpha.
\label{lincorr_eq}
\end{equation}

\begin{figure}
\plotone{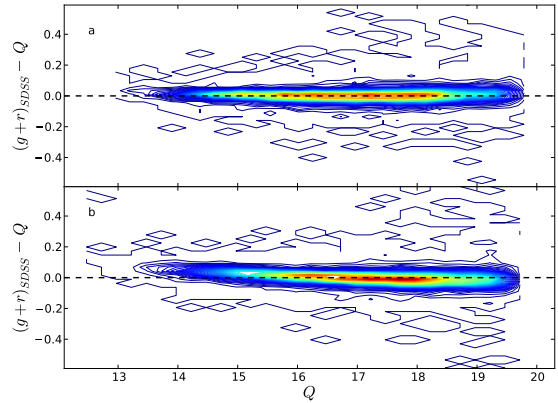}
\caption{Contour level lines of the number of stars as a function of the residual magnitude, and the $Q$-band magnitudes for more than $15,000$ stars.
The residual is the difference between the $(g+r)_{SDSS}$ magnitude and the mean $Q$-band magnitude of the stars. These stars were flagged as
standards based on our 2011 observations. In panel (a) the $Q$-band calibration was done using our linearity correction technique (see Section \ref{lincorr_sec}),
whilst in panel (b) the $Q$-band calibration was done using a constant zero point.}
\label{gprvsQ_phot_fig}
\end{figure}

In Figure \ref{gprvsQ_phot_fig} (top-panel) we show the residuals of $Q$ minus the $(g+r)_{SDSS}$ magnitude for $18,265$ non-variable
stars where the calibration of the $Q$ magnitudes was done following the prescription described above. These stars were classified as
standards following the same criteria of Section \ref{internal_astrometry}, namely $n$~\textgreater~$10$, a standard deviation in its
light curve smaller than the median error observed in the photometry, a $\sigma_{LC}$ \textless~$0.035$ mag, and $\chi^{2}_{LC}$ \textless~$1.5$.
As shown in the top-panel of Figure \ref{gprvsQ_phot_fig} we obtained a good correspondence between $(g+r)_{SDSS}$ and $Q$ magnitudes,
with a mean residual of $0.004 \pm 0.048$ magnitudes.  

If instead we had used a constant zero point to calibrate the observations, where we used $\sigma$-clipping to clean for outliers,
we would have obtained $16,977$ standard stars with the $(g+r)_{SDSS} - Q$ residuals shown in the bottom-panel of Figure \ref{gprvsQ_phot_fig}. 
The correspondence between $(g+r)_{SDSS}$ and $Q$ is good with a mean residual of $0.001 \pm 0.052$.

\begin{figure}
\plotone{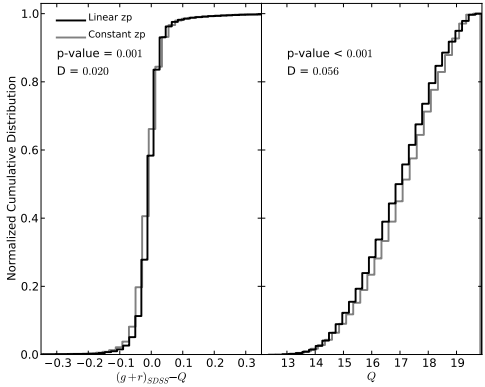}
\caption{Left panel: normalized cumulative distributions of $Q-(g+r)_{SDSS}$ residuals for non-variable point sources for the case of a constant zero--point
(grey), and for a linear model calibration (black). Right-panel: normalized cumulative distributions of $Q$ magnitude for non-variable point sources for the
case of a constant zero--point (grey), and for a linear model calibration (black). The p--value and the Kolmogorov--Smirnov D parameter are indicated on
the left of each panel.}
\label{phot_cumdist_fig}
\end{figure}

Using a linear fit instead of the usual constant zero-point reduces the dispersion in the light curves, thus increasing by $\sim 7.5$\%
the number of non-variables or standard stars. Although the means and the standard deviations of the $(g+r)_{SDSS} - Q$ distributions 
in the case of a linear and constant zero-point are consistent with each other, a Kolmogorov--Smirnov (K--S) test yields a p-value
equal to 0.01. This means that the null hypothesis that both distributions are drawn from the same parent population can be rejected 
(see Figure \ref{phot_cumdist_fig}). Additionally, a K--S test on the $Q$ magnitude distributions yields a p-value lower
than 0.01, so the null hypothesis that both $Q$ magnitude distributions are drawn from the same parent population can be rejected.

We found that when computing the zero-point using a linear model with Eqs.~\ref{res_eq} and \ref{lincorr_eq}, the number of non-variables 
or standard stars increases significantly in the range $15$ \textless $Q$ \textless $19$ (Figure \ref{phot_cumdist_fig}). At the same 
time the linear model keeps the mean and the standard deviation of the $(g+r)_{SDSS} - Q$ residual consistent with the case
of a constant zero-point, and reduces the skewness in $(g+r)_{SDSS} - Q$ as function of $Q$ (see Figure \ref{gprvsQ_phot_fig}).

In summary, we found that the use of a linear model to compute the zero-point as a function of $q$ achieves a better calibration 
of our data, and in the cases where non-linearity is unimportant (i.e., small $\beta$) the photometry remains consistent with
the case of a constant zero-point.

\subsection{Color-term}
\label{color_sec}

The agreement between between the $(g+r)_{SDSS}$ and the $Q$-band is overall very good. However, we decided to explore the possibility of 
adding a color-term to enhance our calibration.

\begin{figure}
\plotone{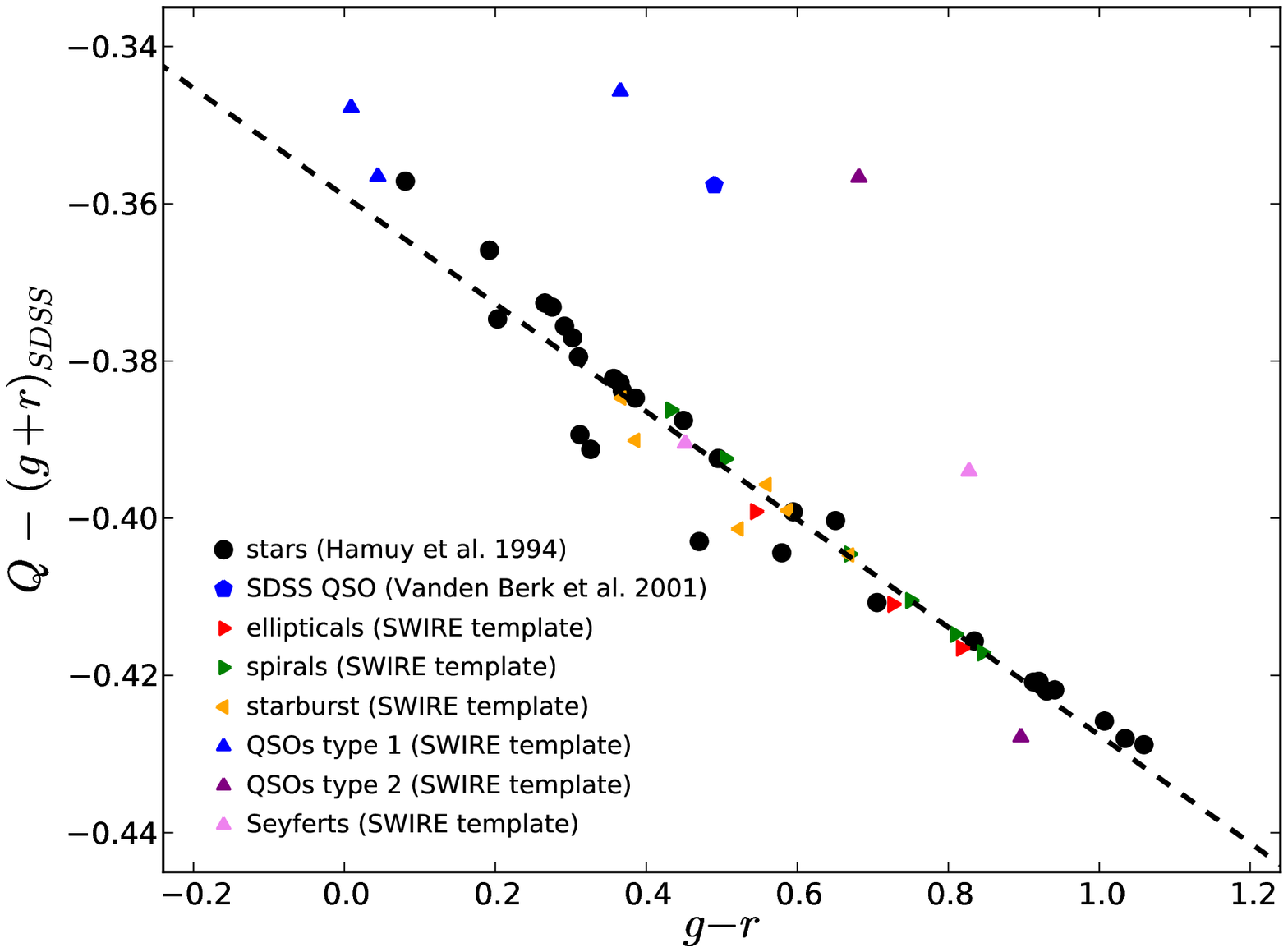}
\caption{$Q - (g+r)_{SDSS}$ residuals of the synthetic photometry versus the $(g-r)_{SDSS}$ synthetic colors of the spectrophotometric standard stars of
\citet{hamuy94, hamuy92}. The black line is the best fit to the stars. Additionally, for comparison we show the $Q - (g+r)_{SDSS}$ residuals of the
synthetic photometry versus $(g-r)_{SDSS}$ synthetic colors of the composite quasar spectra from SDSS \citep{vandenberk01}, in addition to AGNs, galaxies
and starburst galaxies templates from SWIRE \citep[see ][]{polletta07}.}
\label{colorterm_fig}
\end{figure}

With an estimation of the $Q$-band system response at hand, we looked for a relation between the residuals of $Q - (g+r)_{SDSS}$ 
as a function of the $(g-r)_{SDSS}$ color. We began by multiplying the spectrophotometric standard stars of \citet{hamuy94, hamuy92}
with the system response curves of Figure \ref{qfilter_fig}. Then we computed $Q - (g+r)_{SDSS}$, and the $(g-r)_{SDSS}$ synthetic colors.
The results of the synthetic photometry of \citet{hamuy94, hamuy92} standards are shown as black circles in Figure \ref{colorterm_fig}.
We modeled the $Q - (g+r)_{SDSS}$ residuals as a function of the $(g-r)_{SDSS}$ colors using a linear relation (dashed line in Figure \ref{colorterm_fig})
of the form:

\begin{equation}
Q~-~(g+r)_{SDSS}~= -0.359(\pm 0.001)~-0.0686(\pm 0.0014) \times (g-r)_{SDSS}.
\label{colorterm_corr_eq}
\end{equation}

\noindent To estimate the uncertainty in the parameters of the fit, we selected randomly 20 of the 29 stars and then we re-fitted 
a linear model. We repeated this random selection 50 times, then we estimated the standard deviation of the adjusted parameters.

We also obtained synthetic photometry of extragalactic sources to investigate if they follow a relation similar to that of the stars. In Figure 
\ref{colorterm_fig} we show synthetic photometry obtained from SWIRE templates of AGNs, elliptical galaxies, spiral galaxies, 
and starburst galaxies \citep[triangles; see][]{polletta07}. We additionally show synthetic photometry obtained from a SDSS 
composite quasar spectrum \citep[blue pentagon;][]{vandenberk01}. Overall, the template of extragalactic sources follows a linear 
relation similar to that of the stars. The outliers are AGNs, and they depart from the main relation mostly due to the presence of strong 
emission lines.

We used Eq.~\ref{colorterm_corr_eq} to transform the photometry from the $(g+r)_{SDSS}$ photometric system to the $Q$-band. Then
we tested our transformation performing differential photometry on several nights of 2011. We found that defining the $Q$-band as in 
Eq.~\ref{colorterm_corr_eq} does not produce better results than defining the $Q$-band as equivalent to the $(g+r)_{SDSS}$. Indeed using 
Eq.~\ref{colorterm_corr_eq} to compute the $Q$-band magnitude for the reference stars and then to compute the zero points, increases the dispersion 
in the zero point calculation and overall produces worse fits.

\begin{figure}
\plotone{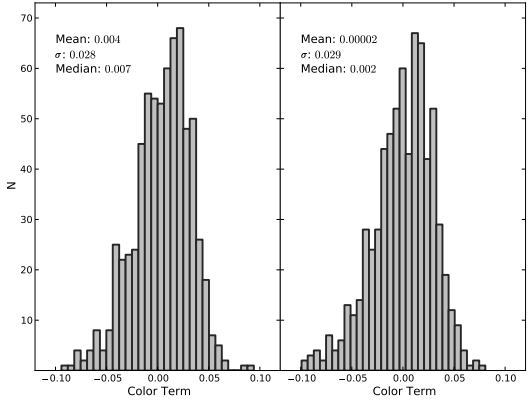}
\caption{Distribution of the color-terms computed from observations of four nights made on 2011. (Left panel) color-term calculated from a $Q$-band
catalog of stars computed assuming a equivalence between the $Q$-band and the $(g+r)_{SDSS}$-band (i.e. zero color-term) but not taking into
account the uncertainty in the color-term correction. (Right panel) color--term calculated from a $Q$-band catalog stars computed
assuming a equivalence between the $Q$-band and the $(g+r)_{SDSS}$-band but taking into account the uncertainty in the color-term correction (see text).
The mean, the median and the standard deviation of the distributions are given on the top-left of each panel.}
\label{colorterm_histo_fig}
\end{figure}

As a second check on the effects of adding a color-term, we performed differential photometry
for reference stars on four clear nights of 2011. We then computed the observed color-term fitting
a straight line to the difference between the instrumental magnitude ($q$) minus the reference $(g+r)_{SDSS}$ magnitudes as a function 
of the $(g-r)_{SDSS}$ colors. In Figure \ref{colorterm_histo_fig} (left panel) we show the distribution of the observed color-terms obtained. 
The mean of the distribution is $0.004$, the median $0.007$, and the standard deviation $0.028$. These values are consistent 
with no color-term. However, the large dispersion in the color-term distribution may be explained 
by a color-term variation between CCDs. Small variations from night-to-night are also expected.

Formally the difference between the computed color-term using the system responses and the observed color-term is $2.59~\sigma$. The 
discrepancy is most likely due to differences between the estimated and the real $Q$-band system responses. For example, we made a
rough assumption of the optics transmission of the camera, and almost certainly there are differences in the quantum efficiency from 
detector-to-detector as function of wavelength. 

In conclusion we find that observations indicate that the color-term is close to zero, and therefore the $Q$-band is well described
by the $(g+r)_{SDSS}$ system. As result we did not apply any color-term to our transformation. Nevertheless, there is some uncertainty associated 
to this assertion. In order to correctly take into account this uncertainty we added in quadrature a color-term 
uncertainty equal to $0.03 \times (g-r)_{SDSS}$. 

In Figure \ref{colorterm_histo_fig} (right panel) we show the resultant distribution of the observed color-terms after adding this color-term
uncertainty. The mean of the distribution was $0.00002$, the median was $0.002$, and the standard deviation obtained
was $0.029$. Overall this additional term did not increase the dispersion in the zero point calculation and produced consistent fits. 
Therefore we assume that it is a robust measure of our uncertainty on the filter transmission discrepancies.

\subsection{Light Curve Generation}
\label{lc_generation_sec}

For each epoch and each detector we generated a catalog which contains the Equatorial position of the object in degrees
(i.e. right ascension and declination), the calibrated magnitude, the magnitude error, 
the number of stars used to obtain the photometric zero-point, the standard deviation of the zero-point, and the reduced $\chi^2$ of 
the zero-point adjustment (see Section \ref{lincorr_sec}). Light curves are generated for objects in the matching catalogs
of different epochs, but always for the same detector, and the same tile. This means that we generate light curves 
for each tile/detector separately. To match sources we used a radius of $1\arcsec$, which is roughly equivalent to 1 pixel. 
We avoid contamination from bad nights, which typically have few stars in the zero-point computation and produce 
a very uncertain photometric zero-point. To this effect, we select only those nights that use more than $20$ stars to compute
the zero-points, where $\chi^2_{\nu}$ \textless~$10$, and the standard deviation of the zero-point adjustment is lower than
$0.10$ mag.

\begin{deluxetable} {lcccc}
\tablecolumns{5}
\tablenum{3}
\tablewidth{0pc}
\tablecaption{Magnitude Difference of the Overlapping Stars.\label{Syserr_tab}}

\tablehead{
\colhead{Range} &
\colhead{N$_{Stars}$} &
\colhead{\textless$\Delta Q$\textgreater} &
\colhead{$\sigma$} &
\colhead{$\Delta Q_{Median}$}}
\startdata

$Q$ \textless $14.0$                  & 40   & -0.044 & 0.065 & -0.036 \\
$14.0$ \textless $Q$ \textless $15.5$ & 316  & -0.020 & 0.056 & -0.016 \\
$15.5$ \textless $Q$ \textless $17.0$ & 1501 & -0.019 & 0.068 & -0.015 \\
$17.0$ \textless $Q$ \textless $18.5$ & 2844 & -0.015 & 0.067 & -0.013 \\
$Q$ \textgreater 18.5                 & 1346 & -0.014 & 0.073 & -0.014 \\
$15.5$ \textless $Q$ \textless $18.5$ & 4345 & -0.016 & 0.067 & -0.014 \\
\hline
Total                                 & 6047 & -0.016 & 0.068 & -0.014 \\

\enddata
\end{deluxetable}

\subsection{Systematic Error in the Photometric Zero-Point}

\begin{figure}
\plotone{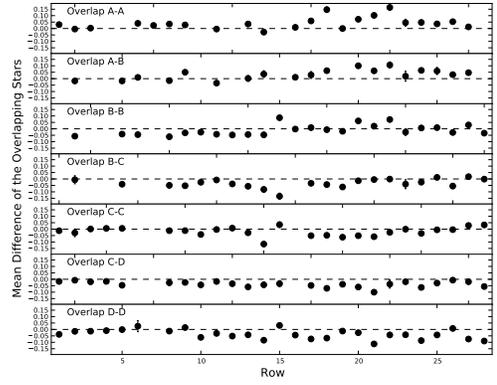}
\caption{Mean differences for the stars in common for each overlapping area. The corresponding overlapping area is indicated in the top-left of each panel.}
\label{Syserr_fig}
\end{figure}

We assessed systematic differences in our photometry by comparing the magnitudes of stars in the common area of 
two adjacent tiles offset by $0\fdg5$. We found that there are $6047$ stars detected in two 
adjacent images of the COSMOS field. The mean difference in their magnitudes is $0.016 \pm 0.068$ magnitudes, with
a median difference of $-0.014$ magnitudes. In Figure \ref{Syserr_fig} we summarize the mean magnitude differences
for each overlapping area. We investigated the magnitude differences as a function of magnitude dividing the
stars in magnitudes ranges, and we found that stars in the brighter range have higher systematic differences than 
fainter stars. Our results are summarized in Table \ref{Syserr_tab}. 

Based on our results we quote that our systematic uncertainty is $0.05$ magnitudes, this value brings into 
agreement most of our observed/calibrated photometry in different tiles. We note that this systematic 
uncertainty is related to variations from detector-to-detector, and that additional variations are also 
introduced as a consequence of the use of different stars to compute the zero-points. For observations 
obtained using the same tile/detector it is not necessary to take in to account this uncertainty, since we are using
the same detector and mostly the same stars to compute the zero point.

\section{OPTICAL VARIABILITY OF XMM-COSMOS X-RAY SELECTED AGNs}
\label{op_var_vs_multiwave}

In this section we study the optical variability properties of the XMM-COSMOS X-ray selected AGNs. To do 
this we constructed optical light curves using the 2010--2012 data for 287 QUEST detected sources out of the 1797
X-ray point sources presented in Table 2 of \citet{brusa10} ({\it the X-ray catalog\/} hereafter). Most of the
objects contained in the {\it the X-ray catalog\/} are faint optical sources with a mean $r$ magnitude of
$\sim 22.9$ and a standard deviation of $1.9$ mag in their $r$ magnitude distribution. Overall, we detected
$\sim 76$\% (254 objects) of the sources in the {\it the X-ray catalog\/} to a limiting magnitude of $r \sim 21$.
{\it The X-ray catalog\/} contains X-ray, optical, infrared, spectroscopic, and photometric redshift information
for the XMM-COSMOS sources. To construct this multiwavelength data set \citet{brusa10} cross-correlated
X-ray positions with the optical multiband catalog of \citet{capak07}, the Canada--France--Hawaii Telescope
(CFHT) K-band catalog \citep{mccracken10}, the IRAC catalog \citep{sanders07, ilbert09}, and the $24$ $\mu$m MIPS
catalog \citep{lefloch09}. Additionally, they used accurate C-Chandra \citep{elvis09} positions available
for a subset of objects within an area of $\sim 1$ deg$^{2}$ to control-check the optical/near-IR identifications,
and to assess the reliability of the proposed identifications.

Good-quality spectroscopic redshifts for the optical counterparts are available from Magellan/IMACS and 
MMT observation campaigns \citep[$\sim 530$ objects;][]{trump07, trump09}, from the VIMOS/zCOSMOS 
\citep[$\sim 580$ objects;][]{lilly07, lilly09}, or were already present either in the SDSS catalog 
\citep[$\sim 100$ objects;][]{adelman-mccarthy06, kauffmann03}, or in the literature \citep[$\sim 95$ objects;][]{prescott06}.
Using this large spectroscopic data set, \citet{brusa10} divided the extragalactic sources with available spectra 
into three classes, on the basis of a combined X-ray and optical spectroscopic classification.

\begin{list}{$\bullet$}{}

\item {\it Broad-line AGN (BL AGN)}: all objects having at least one broad 
(FWHM \textgreater $2000$~km~s$^{-1}$) optical emission line in the available spectrum; (421 sources).

\item {\it Non-broad-line AGN (NL AGN)}: all objects with unresolved, high-ionization
emission lines, exhibiting line ratios indicative of AGN activity; if lines are not detected or the observed spectral range does not allow to construct line 
diagnostics, objects with rest-frame hard
X-ray luminosity in excess of $2 \times 10^{42}$~erg~s$^{-1}$, typical of AGNs; (370 sources).

\item {\it ``Normal'' galaxies (GALs for short)}: all sources with unresolved emission lines consistent with 
spectra of star-forming galaxies or galaxies showing only absorption lines, and with 
rest-frame hard X-ray luminosity lower than $2 \times 10^{42}$~erg~s$^{-1}$, or undetected in the 
hard band; (53 sources).

\end{list}

\subsubsection{$u-g$ color vs. $g-r$ color}

\begin{figure}
\plotone{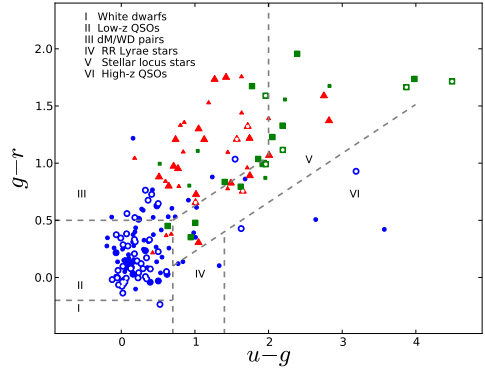}
\caption{
$u-g$ vs. $g-r$ color-color diagram for all the XSMM-COSMOS extragalactic sources detected in the QUEST-La Silla survey. Blue points
correspond to BL AGNs, red triangles to NL objects, and green squares to GAL sources. Larger symbols are used to highlight the best
sample, and white spot identify variable sources (see Section \ref{variability}). To guide the eye we show the division used in
\citet{sesar07} as grey dashed lines, that it corresponds to regions where different subclasses of variables objects are dominant.
The classes are listed on the top-left of the figure. The photometry correspond to model magnitudes obtained from the SDSS-DR7 database,
using a cross-matching radius of $0\farcs3$ to match objects. The photometry was corrected using Galactic extinction values given by SDSS,
that make use of the extinction maps of \citet{schlegel98}.
}
\label{u_g_g_r_brusa10_fig}
\end{figure}

In Figure \ref{u_g_g_r_brusa10_fig} we show the $u-g$ vs. $g-r$ color-color diagram for all the XMM-COSMOS extragalactic sources detected in 
our survey constructed using SDSS-DR7 photometry \citep{abazajian09}. In the figure blue circles correspond to BL AGNs, red triangles
to NL AGNs, and green squares to GALs. We use larger symbols to highlight the best observed sample defined below in Section \ref{variability},
and a white spot to identify variable sources (see Section \ref{variability}). To guide the reader's eye we plot 
the division used in \citet{sesar07} as grey dashed lines, which corresponds to regions where different subclasses of variables 
objects are dominant. For example, in regions II and VI the dominant classes of variable objects are low-z and high-z QSOs, 
respectively.

The photometry corresponds to model magnitudes obtained from the SDSS-DR7 database \citep{abazajian09}, using a cross-matching radius
of $0\farcs3$ to match objects. Using this cross-matching radius we find that 97\% of BL AGNs, 88\% of NL AGNs, and 83\% of GALs have
a SDSS counterpart. The photometry was corrected using Galactic extinction values given by SDSS, that make use of the extinction 
maps of \citet{schlegel98}. 

From the figure it is clear that in the $u-g$ vs. $g-r$ color-color space BL AGNs, NL AGNs, and GALs occupy roughly different regions.
While BL AGNs tend to be concentrated in region II (low-z QSOs), with a few objects in region VI (high-z QSOs), and a non-negligible
fraction of objects in region III (dM/WD pairs); NL AGNs are concentrated in region III, with some objects in region V 
(Stellar locus stars), and very few objects in region II. On the other hand, GALs are concentrated in region V, with a non-negligible
fraction of objects in region III. In particular, GAL objects seems to occupy the color-color region between region III and region V.

As expected, BL AGN colors are dominated by AGN emission, while GAL colors are dominated by stellar emission. On the other hand, the 
optical colors of NL AGNs are consistent with obscured AGNs with some contribution from the host galaxy stellar emission. Since
\citet{brusa10} classification is partly based on optical spectra this confirms that they classification is robust and consistent
with the optical colors.

\subsection{QUEST-La Silla Variability Survey Completeness and Detection Limit}

Of the $287$ XMM-COSMOS sources detected by QUEST, $233$ sources have a corresponding $r$-band magnitude in the {\it X-ray catalog}.
The remaining $44$ sources are bright, probably saturated, and therefore not included in the optical 
multiband catalog. In Table \ref{Comple_tab} we summarize the detection completeness of our survey as function
of $r$ magnitude ranges for easy comparison with other surveys, we also breakdown our analysis for different classes 
of extragalactic objects. Table \ref{Comple_tab} is organized as follows: $r$ magnitude range (Column 1); total number 
of sources detected in our survey in the corresponding magnitude range (Column 2); number of objects detected in our survey
over the total number of XMM-COSMOS objects in the corresponding magnitude range (Column 3); number of BL AGNs detected in
our survey over the total number of XMM-COSMOS BL AGNs in the same magnitude range (Column 4); number of BL AGNs detected
in our survey over the total number of sources detected in our survey; in parentheses the number of total XMM-COSMOS BL AGNs
over the total number of XMM-COSMOS sources in the magnitude range (Column 5); number of NL AGNs detected over the XMM-COSMOS
NL AGNs in the corresponding magnitude range (Column 6); number of NL AGNs detected over the total number of sources detected
in the range, and in parentheses the number of XMM-COSMOS NL AGNs over the total number of XMM-COSMOS sources in the range (Column 7);
number of GALs detected over the total number of XMM-COSMOS GALs in the range (Column 8); number of GALs detected over 
the total number of sources detected in our survey, and in parentheses the number of total XMM-COSMOS 
GALs over the total number of XMM-COSMOS sources in the range (Column 9).

\begin{deluxetable*} {ccccccccc}
\tablecolumns{9}
\tablenum{4}
\tablewidth{0pc}
\tablecaption{QUEST-La Silla AGN Variability Survey Completeness in the COSMOS Field.\label{Comple_tab}}

\tablehead{
\colhead{Range} &
\colhead{n$_{det}$} &
\colhead{n$_{det}$/n$_{tot}$} &
\colhead{n$^{bl}_{det}$/n$^{bl}_{tot}$} &
\colhead{n$^{bl}_{det}$/n$_{det}$ (n$^{bl}_{tot}$/n$_{tot}$)} &
\colhead{n$^{nl}_{det}$/n$^{nl}_{tot}$} &
\colhead{n$^{nl}_{det}$/n$_{det}$ (n$^{nl}_{tot}$/n$_{tot}$)} &
\colhead{n$^{ga}_{det}$/n$^{ga}_{tot}$} &
\colhead{n$^{ga}_{det}$/n$_{det}$ (n$^{ga}_{tot}$/n$_{tot}$)}\\ 
\hline
\colhead{$1$} &
\colhead{$2$} &
\colhead{$3$} &
\colhead{$4$} &
\colhead{$5$} &
\colhead{$6$} &
\colhead{$7$} &
\colhead{$8$} &
\colhead{$9$}
}

\startdata


$17$ \textless $r$ \textless $18$ & $9$   & $0.75$           & \nodata   & $0.00$($0.00$) & $1.00$           & $0.33$($0.25$) & $0.62$ & $0.56$($0.67$) \\ 
$18$ \textless $r$ \textless $19$ & $26$  & $0.87$           & $0.80$    & $0.31$($0.33$) & $0.80$           & $0.15$($0.17$) & $0.88$ & $0.27$($0.27$) \\ 
$19$ \textless $r$ \textless $20$ & $58$  & $0.74$           & $0.72$    & $0.45$($0.46$) & $0.73$           & $0.19$($0.19$) & $0.75$ & $0.16$($0.15$) \\
$20$ \textless $r$ \textless $21$ & $107$ & $0.67$           & $0.74$    & $0.62$($0.57$) & $0.59$           & $0.16$($0.18$) & $0.60$ & $0.06$($0.06$) \\
$21$ \textless $r$ \textless $22$ & $30$  & $0.12$           & $0.20$    & $0.67$($0.40$) & $0.06$           & $0.17$($0.32$) & $0.00$ & $0.00$($0.04$) \\
$22$ \textless $r$ \textless $23$ & $3$   & $0.01$\textless  & $0.02$    & $0.67$($0.33$) & $0.01$\textless  & $0.33$($0.33$) & $0.00$ & $0.00$($0.01$\textless) \\

\enddata
\end{deluxetable*}

The survey is roughly $75$\% complete in the magnitude range between $r \simeq 17$ mag and $r \simeq 20$ mag,
and slightly less complete ($67$\%) up to $r \simeq 21$ mag. In the range between $r \simeq 21$ and $r \simeq 22$ mag 
the survey declines to $12$\% completeness, and the survey drops to less than $1$\% completeness at fainter magnitudes. 
If we break down the detection completeness fraction for different object classes ($n^{class}_{det} / n^{class}_{tot}$; 
see Table \ref{Comple_tab}), we find that the detection completeness fraction, for all classes of objects, is consistent 
with roughly $75$\% up to $r \simeq 20$ mag. In the $r$ magnitude range between $20$ mag and $21$ mag the detection fraction
of BL AGNs is $74$\%, whilst the detection fraction of NL AGNs and GALs is $60$\%. At fainter magnitudes the
detection fraction of BL AGNs is significantly higher than that of NL AGNs and GALs. This could be due to strong UV rest-frame
emission lines (Ly-$\alpha$, Si~IV, C~IV, and C~III) entering the blue side of the Q-band for sources at z \textgreater $2.0$.
When we break down the number of objects detected of a certain class over the total number of objects detected in a magnitude 
range ($n^{class}_{det} / n_{det}$; see Table \ref{Comple_tab}), they are in very good agreement with the number 
of XMM-COSMOS objects of the same class over the total number of XMM-COSMOS objects in the sample 
($n^{class}_{tot} / n_{tot}$; see Table \ref{Comple_tab}) in the magnitude range up to $r \simeq 21$.

The analysis of the detection fractions indicates that we are detecting a large fraction of objects up to $r \simeq 21$.
This is in agreement with the limiting magnitude expected based on the exposure time used; for an exposure time of 60 seconds 
the expected limiting magnitude is $Q \sim 19.5$ which is equivalent to a $r$ magnitude in the range between $20$ mag and
$21$ mag. Similarly, for an exposure time of 180 seconds the expected limiting magnitude is $Q \sim 20.5$ equivalent to a
$r$ magnitude in the range between $21$ mag and $22$ mag. During 2011 and 2012 most of our observations were performed using
exposure times of 60 seconds, where the completeness below $r \sim 21$ mag is low. Currently (2013-2014) we are performing our 
observations using exposure times of $180$ seconds which will increase our completeness at fainter magnitudes.

\begin{figure}
\plotone{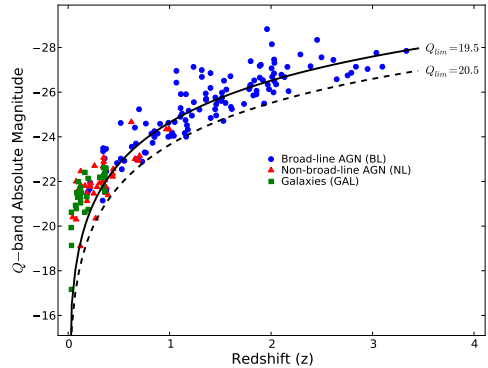}
\caption{Q-band absolute magnitudes for XMM-COSMOS sources detected as part of our survey. Absolute magnitudes are not
K-corrected or corrected by dust extinction. Blue points correspond to broad-line AGN, red triangles correspond to objects
classified as non-broad-line AGN, and green squares are classified as ``normal'' galaxies or starburst galaxies.
The continuous line is the detection limit for an object with $Q=19.5$ magnitude, and the dashed line is the
detection limit for an object with $Q=20.5$ magnitude.}
\label{detection_lim_fig}
\end{figure}

To illustrate our detection limit as function of redshift, in Figure \ref{detection_lim_fig} we show the $Q$-band
absolute magnitude of the extragalactic XMM-COSMOS objects as function of redshift. The $Q$-band absolute magnitudes are 
not K-corrected or corrected by dust extinction. As can be seen we are detecting objects up to redshift $\sim 3$. In
total we detect $29$ GALs all below redshift $0.4$, $42$ NL AGNs all below redshift $1.0$, and 123 BL AGNs. In Figure
\ref{detection_lim_fig} we additionally show the detection limit for an object with $Q = 19.5$ mag (continuous line),
and the detection limit for an object with $Q = 20.5$ mag (dashed line).

\subsection{Variability of the XMM-COSMOS Objects}
\label{variability}

To study the variability of the XMM-COSMOS extragalactic sources we defined a flux limited sample ($Q$ \textless 19.5), 
where the time lapse between the first and the last observations was more than $600$ days. We call this the {\it best observed\/} sample. 
The time span ensures that the light curves have a significant number of observations. Additionally the time span longer 
than 600 days ensures that we will be able to study long-term variability as this corresponds to rest-frame timescales 
of at least a few months for objects at $z$~\textgreater~$0.6$, and of $\ga 1$ year for objects at
$z$~\textless~$0.6$. Our sample of best observed light curves consists of $55$ BL AGNs, $24$ NL AGNs, and $23$ GALs. 

To classify an object as variable or non-variable we used the variability index V \citep{mclaughlin96, paolillo04, lanzuisi14}, 
defined as $V= -$log$(1 - P(\chi^{2}))$, where $P(\chi^{2})$ is the the probability that a $\chi^{2}$ lower or equal to the observed 
value could occur by chance for an intrinsically non-variable source. We also defined $P_{var} = 1 - P(\chi^{2})$ as the probability
of an object of being variable. Following \citet{lanzuisi14} we defined $V=1.3$ as our threshold to define a variable object,
since an object with $V$ \textgreater $1.3$ has more than $95$\% probability to be variable ($P_{var}$ \textgreater $0.95$).
Based on the V index we find that 44 ($80 \pm 5$\%) of our best observed BL AGNs are variable, 5 ($21 \pm 8$\%) of our best observed
NL AGNs are variable, and 5 ($22 \pm 9$\%) of our best observed GALs are variable. Since the division between variable and non-variable
objects is somewhat arbitrary, we assumed a binomial distribution in the classification to estimate the errors. Therefore, based on
the binomial distribution 3 BL AGNs, 2 NL AGNs, and 2 GALs may be classified as variable sources by chance.

As a measure of the variability amplitude we used the excess variance defined as $\sigma^2_{rms} = \sigma^{2}_{LC} - \sigma^{2}_{median}$,
where $\sigma_{LC}$ is the standard deviation of the light curve, and $\sigma_{median}$ is the median error of the light curve.
In the case where $\sigma^{2}_{LC}$ \textless $\sigma^{2}_{median}$ we used the median error as an upper limit on the excess
variance. In Table \ref{ExcessVar_tab} we present the distribution of $\sigma_{rms}$ for our sample of best observed objects
divided in three bins ($\sigma_{rms} < 0.05$, $0.05 < \sigma_{rms} < 0.10$, and $\sigma_{rms} > 0.10$).
We found that BL objects tend to have larger variability amplitude, with $77$\% of the objects showing $\sigma_{rms} > 0.05$,
and $23$\% of the objects with $\sigma_{rms} > 0.10$. On the other hand, $33$\% of the NL AGNs and GALs objects have
$\sigma_{rms} > 0.05$. Of these, 1 NL AGN ($17$\%) have $\sigma_{rms} > 0.10$ while all GALs have
$\sigma_{rms}$ below this value. When we include upper limits to the distributions the tendency remains the same with $80$\% of BL AGNs,
$58$\% of NL AGNs, and $44$\% of GALs having $\sigma_{rms} > 0.05$.

\begin{deluxetable} {lccc}
\tablecolumns{4}
\tablenum{5}
\tablewidth{0pc}
\tablecaption{Excess Variance Distribution.\label{ExcessVar_tab}}

\tablehead{
\colhead{Range} &
\colhead{BL} &
\colhead{NL} &
\colhead{GAL}}
\startdata

$\sigma_{rms}$ \textless $0.05$                  & $0.23$ ($0.20$) & $0.67$ ($0.42$) & $0.67$ ($0.56$) \\
$0.05$ \textless $\sigma_{rms}$ \textless $0.10$ & $0.54$ ($0.56$) & $0.17$ ($0.50$) & $0.33$ ($0.35$) \\
$\sigma_{rms}$ \textgreater $0.10$               & $0.23$ ($0.24$) & $0.17$ ($0.08$) & $0.00$ ($0.09$) \\

\enddata
\tablecomments{Fraction of BL, NL, and GAL objects in a given range of $\sigma_{rms}$ for the best observed sample.
In parentheses we give the fraction of objects in a given range, but now including
upper limits.}
\end{deluxetable}

\subsubsection{Excess Variance vs. $[3.6] - [8.0]$ Color}

The mid-IR colors have been proposed as a selection method to identify AGNs \citep[see][and references therein]{lacy04, stern05, kozlowski09, assef13}.
The method relies on the fact that the mid-IR emission of normal galaxies and AGNs have different spectral energy 
distributions (SED). The composite spectra of the stellar population of normal galaxies produces a SED that peaks
at approximately $1.6$ $\mu$m, while AGNs have a roughly power-law SED due to hot dust emission, that reprocess
the UV/optical emission of the accretion disk, peaking somewhere around $10$~$\mu$m \citep{mor12, lira13}. 
Following this idea \citet{brusa10} showed that the IRAC broadband $[3.6] - [8.0]$ mid-IR color 
can be used to disentangle objects with infrared rising SED (red power-law), e.g. AGNs, and sources 
with an inverted SED (blue power-law), e.g., normal galaxies at low-z. They used this method to 
select a sample of highly obscured luminous AGN candidates.

\begin{figure}
\plotone{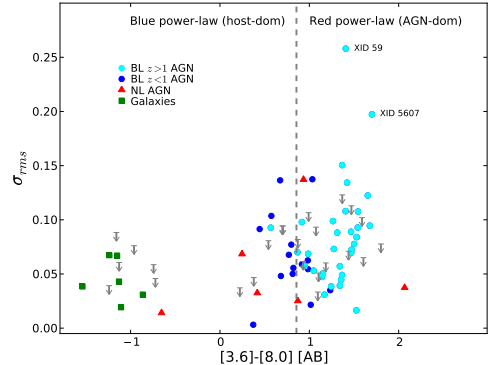}
\caption{
$\sigma_{rms}$ vs. $[3.6] - [8.0]$ color for the sample of best observed XMM-COSMOS objects. In the figure cyan cirlces are BL AGNs
with $z$ \textgreater $1$, BL AGNs with $z$ \textless $1$ are shown as blue circles,
red triangles correspond to NL AGNs, and green squares are GALs. The grey dashed line corresponds to the division between
AGN dominated and host dominated sources of \citet{brusa10}, and upper limits are shown as grey arrows down for BL and NL AGNs,
and not shown for GALs for clarity.}
\label{midir_vs_sigma_fig}
\end{figure}

In Figure \ref{midir_vs_sigma_fig} we explore a tentative relation between the $\sigma_{rms}$ and the $[3.6] - [8.0]$ 
mid-IR color for our sample of best observed BL AGNs, NL AGNs, and GALs. In the Figure we show BL AGNs with $z$ \textgreater $1$ 
using cyan circles, BL AGNs with $z$ \textless $1$ using blue circles, red triangles correspond to NL AGNs, and green 
squares are GALs. The grey dashed line corresponds to the division between AGN dominated and host dominated sources of 
\citet{brusa10} ($[3.6] - [8.0] = 0.856$), and upper limits are shown as grey downward arrows for BL and NL AGNs, and not shown
for GALs for clarity.

As can be seen in Figure \ref{midir_vs_sigma_fig} BL AGNs show on average large $\sigma_{rms}$. Among them there is an
apparent trend, although with large scatter, where AGNs with higher $\sigma_{rms}$ tend to have redder $[3.6] - [8.0]$ colors.
This trend is largely driven by BL AGNs, particularly those at $z$ \textgreater $1$ (cyan circles), where the expected contribution
from the host galaxy to the $Q$-band is very low. Among the BL AGNs with $z$ \textgreater $1$ the trend seems not to be related
to redshift or luminosity. Finally, the BL AGNs with $z$ \textless $1$ have bluer colors than the overall BL AGN sample which
implies a higher contribution of host galaxy light to the photometry. 

A plausible interpretation for the trend shown in Figure \ref{midir_vs_sigma_fig} is that AGN variability is diluted
by host galaxy light in objects with bluer mid-IR colors (i.e., more host-dominated).
This will be explored by our collaboration in the future using a larger sample obtained from photometry of stacked images
and with light curves with longer time span.

\subsubsection{Structure Function}

The structure function is a simple tool to quantify the variability of a source as a function of the time lapse
between observations ($\tau$) when a significant number of observations are available. To study the variabilty
behaviour of our sample as a function of time we calculated the structure function as:

\begin{equation}
SF(\tau) = \sqrt{\frac{1}{N} \sum_{i=1}^{N} [m(t_i) - m(t_i + \tau)]^{2}}. \\
\end{equation}

\noindent where $m(t_i)$ is the magnitude at time $t_i$, $m(t_i + \tau)$ is the magnitude
at time $t_i + \tau$ in the rest--frame, and $N$ is the number of observations in the time span bin.
To compute the structure function we average values in equal size bins in logarithmic space. The
bins are "centered'' in $2^{i}$ days, where the bin interval is $[2^{i-\frac{1}{2}}, 2^{i+\frac{1}{2}}]$
with $i=0,1,2,3,...$

To take into account the errors in the measurements we generated a random number drawn from a
Gaussian distribution with mean and standard deviation equal to the measurement value and its
uncertainty, respectively. Then, we computed the structure function using this mock data set
and we ``normalized'' it removing in quadrature the contribution produced by the signal variance
estimated as the median value of the uncertainties in the measurements ($\sigma_{median}$).

As it is shown in \citet{schmidt10} and \citet{palanque11} the structure function of a QSO
is well described by a power-law, where $\tau$ is measured in years, as:

\begin{equation}
SF_{norm}(\tau) = A \tau^{\gamma} \\
\end{equation}

In Figure \ref{sf_param_fig} we show the power-law exponent ($\gamma$) and amplitude ($A$) that resulted
from the power-law fit to our sample. Note that $A=SF(\tau = 1)$. All the objects classified as variables
based on the V parameter are marked with a white spot on Figure \ref{sf_param_fig}. It is clear from
the figure that the variability of BL AGNs is on average described by larger $A$
and $\gamma$ values compared to NL AGNs and GALs. The latter objects tend to be clustered close to
$\gamma \simeq 0.05$ and $A \simeq 0.05$ in the $\gamma$--$A$ plane. Based on this we defined
the region $\gamma$ \textgreater $0.025$ and $A$ \textgreater $0.1$ in which the observed
variability is consistent with being powered by the accretion disk of
a super-massive black hole.
Therefore, NL AGNs and GALs lying in this region are suspected to show variability consistent with an AGN.

\begin{figure}
\plotone{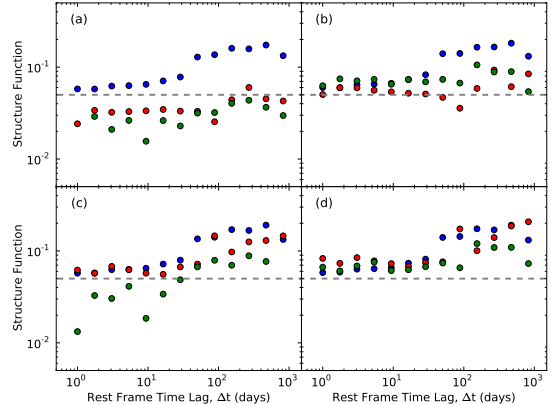}
\caption{Ensemble structure functions for BL AGNs (blue), NL AGNs (red), and GALs (green) for different subsamples, the dashed-grey line corresponds to $0.05$ mag
the limit value from which any variability signal should dominate over significant noise on the QUEST data. (a) Structure functions for the best observed
sample of BL AGNs (n=55), NL AGNs (n=24), and GALs (n=23). (b) Structure functions for the objects of the best observed sample classified as variables
according to $V$ \textgreater $1.3$, namely 44 BL AGNs, 5 NL AGNs, 5 and GALs. (c) Structure functions for the objects of best observed sample
located within parameter space region of Figure \ref{sf_param_fig} dominated by BL AGNs ($A$ \textgreater $0.1$ and $\gamma$ \textgreater $0.025$),
namely 48 BL AGNs, 7 NL AGNs, and 7 GALs. (d) Structure functions for the objects of best observed sample located within parameter space
region of Figure \ref{sf_param_fig} dominated by BL AGNs and classified as variables; 41 BL AGNs, 3 NL AGNs, and 2 GALs.}
\label{sf_compar_fig}
\end{figure}

In Figure \ref{sf_compar_fig} we investigate the connection between the observed variability in NL AGNs and GALs and
the variability powered by the accretion disk of a super-massive black hole represented by BL AGNs. We compare ensemble
structure functions of BL AGNs shown in blue, NL AGNs shown in red, and GALs shown in green when we apply different
selection criteria to define the sample used to compute the ensemble structure functions. 
In panel (a) we show the ensemble structure functions when we consider all objects in the best observed sample; in panel (b) we show
the ensemble structure functions when we consider only variable objects ($V$ \textgreater $1.3$); in panel (c) we show
the ensemble structure functions when we consider objects located in the BL AGN dominated region defined in Figure \ref{sf_param_fig}
($\gamma$ \textgreater $0.025$ and $A$ \textgreater $0.1$); finally, in panel (d) we show the structure functions when
we consider objects that are located in the BL AGN dominated region (see Figure \ref{sf_param_fig})
and are classified as variables according to the $V$ parameter.

In panel (a) can be seen that when we consider all objects in the best observed sample the ensemble structure functions
of the NL AGNs and GALs are nearly flat, and very different from the AGN powered variability of BL AGNs that are characterized
by an increase of the amplitude of the ensemble structure function as function of $\tau$.
On the other hand, clear variability signatures begin to appear in the ensemble structure functions of 
NL AGNs and GALs when we consider variable objects (panel b) or objects located in the BL AGN dominated region in
the $\gamma$--$A$ plane (panel c).
However, only when we restrict our selection criteria to variable objects ($V$ \textgreater $1.3$) located in the BL AGN dominated
region ($\gamma$ \textgreater $0.025$ and $A$ \textgreater $0.1$) we obtain ensemble structure functions for both NL AGNs and GALs
showing variability as function $\tau$ consistent with being powered by an AGN (i.e., similar to BL AGNs; see panel d). Our sample of
variable objects with structure function parameters consistent with AGN powered variability correspond to 74.5\% (41) of BL AGNs,
12.4\% (3) of NL AGNs, and 8.7\% (2) of GALs. This implies that 3/5 and 2/5 of the variable NL AGNs and GAL sources selected based 
on the $V$ index are now securely classified as variable AGNs. This is also consistent with the estimates of false positives derived
using the binomial distribution.

\begin{figure}
\plotone{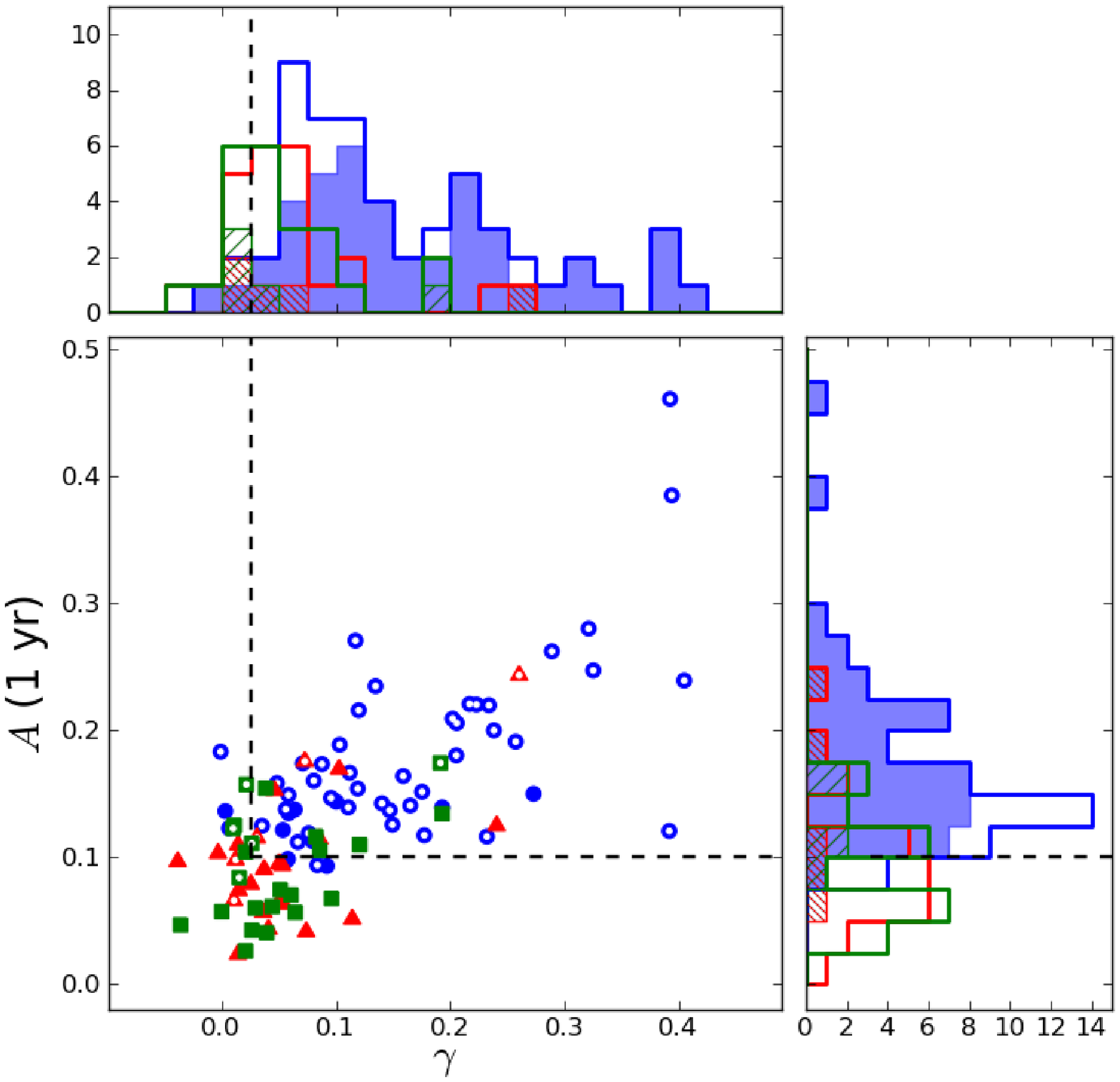}
\caption{Amplitude ($A$) and power-law exponent ($\gamma$) obtained from fitting a power law to the structure
function to the {\it best observed\/} sample of BL AGNs (blue), NL AGNs (red), and GALs (green). We mark with a
white spot the objects classified as variables based on the $V$ parameter (see Section \ref{variability}).
In the top-panel we show the distribution of the $\gamma$ for BL AGNs (blue line), NL AGNs
(red line) and  GALs (green line). In the top-panel the blue shaded distribution corresponds to variable BL AGNs, while the red and green hatched
distributions correspond to variable NL AGNs and GALs, respectively. Similarly, on the right-panel we show the distributions of the amplitude ($A$).
The black dashed line demarcates the region in the $A$--$\gamma$ parameter space dominated by BL AGNs (namely $A$ \textgreater $0.1$ and
$\gamma$ \textgreater $0.025$), and therefore the region where the observed variability can be considered as being powered
by the accretion disk of a supermassive black hole.}
\label{sf_param_fig}
\end{figure}

\subsubsection{Distribution of the $[3.6] - [8.0]$ Color}

\begin{figure}
\plotone{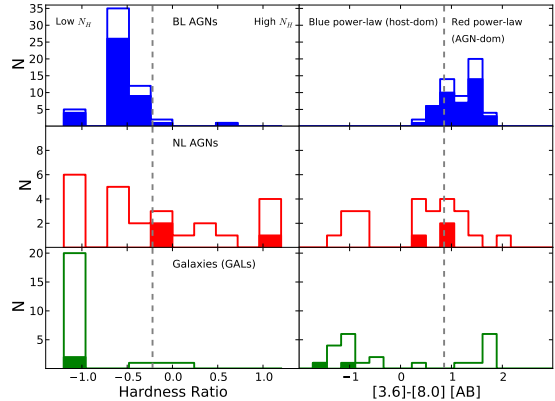}
\caption{
In the left-panel of we show the  overall distribution (step-line histogram) of BL AGNs (top-panel),
NL AGNs (middle-panel), and GAL ojects (bottom-panel) as function of the Hardness Ratio, whilst in the right-panel
we show the same distributions as function of the $[3.6] - [8.0]$ color. The distributions of variable sources
are shown as shaded histograms in the panels. In the left-panel the grey dashed line corresponds to HR=$-0.2$,
the division between unobscured (low N$_H$) and obscured (high N$_H$) AGNs, and in the right-panel the grey dashed
line corresponds to the division between AGN dominated and host dominated sources of Brusa et al. (2010).
}
\label{V_vs_midir_HR_fig}
\end{figure}

On the right-panel of Figure \ref{V_vs_midir_HR_fig} we show the overall distribution (step-line histogram) of BL AGNs
(top-panel), NL AGNs (middle-panel), and GALs (bottom-panel) as function of the $[3.6] - [8.0]$ color. The distributions of
variable sources with $\gamma$ \textgreater $0.025$ and $A$ \textgreater $0.1$ (i.e., $SF(\tau)$ similar to BL AGNs) are shown
in the panels as shaded histograms. The grey dashed line corresponds to the division between AGN dominated and host dominated
sources of \citet{brusa10}.

As expected all BL AGNs show red colors with the majority of them consistent with being dominated by the emission from hot dust.
Variable BL AGNs do not show any trend with the $[3.6] - [8.0]$ color. The $[3.6] - [8.0]$ color distribution of
NL AGNs shows a large spread. There is a group of NL AGNs with very blue colors, and a second group with redder colors peaking on
the blue side of the BL AGN distribution, and lying close to the division between AGN dominated and host dominated sources
of \citet{brusa10}. It is important to mention that, although the NL AGN sample is small, among the objects located on the blue side
of the BL AGN color distribution we find the ones with the larger excess variance (see Figure \ref{midir_vs_sigma_fig}). The GALs
shows a large spread in the $[3.6] - [8.0]$ color distribution. However, all variable sources have very blue colors and are consistent 
with being host-dominated objects.

\subsubsection{Variability vs. Hardness Ratio}

The Hardness Ratio (HR) can be a measure of the X-ray obscuration suffered by an AGN, and it is defined as HR = (H-S)/(H+S),
where S and H are the count rates in the soft ($0.5-2.0$~keV) and hard bands ($2.0-10$~keV). High hydrogen column densities
($N_{H}$) block soft X-rays, hence the effect of increasing column densities is that the X-ray spectrum becomes harder.
Many studies use a $N_{H}$ value of $10^{22}$~cm$^{-2}$ to separate between obscured and unobscured AGNs; \citet{mainieri07}
showed that 90\% of the sources with column densities larger than $10^{22}$~cm$^{-2}$ have HR \textgreater $-0.3$. On the other hand,
\citet{hasinger08} found that using a HR value of $-0.2$ as a threshold can distinguish between obscured and unobscured AGNs.
We will explore the relation between variability and HR using the latter value, also used by \citet{brusa10}, as a threshold. 
Note that the flux limit in the soft band is $\sim 6$ times fainter than in the hard band \citep{brusa10}, and that objects
detected only  in the soft or hard X-ray band have values of HR$=-1$ and HR$=1$, respectively. While most of our GAL objects have 
HR$=-1$, which stems from the \citet{brusa10} criteria to classify X-ray sources in the GAL class (see beginning of Section 7), 
Hasinger's (2008) sample of galaxies shows typical values of $-0.2$\textless~ HR \textless$0.5$ (see his Figure 2). 
This can be explained as \citet{hasinger08} used 2-10 keV detected samples only.

In the left panel of Figure \ref{V_vs_midir_HR_fig} we show the overall distribution (step-line histogram) of BL AGNs 
(top-panel), NL AGNs (middle-panel), and GALs (bottom-panel) as a function of HR. The distributions of variable sources
with $\gamma$ \textgreater $0.025$ and $A$ \textgreater $0.1$ (i.e., $SF(\tau)$ similar to BL AGNs) are shown as shaded
histograms. The grey dashed line corresponds to HR$=-0.2$, the division between unobscured (low $N_{H}$), 
and obscured (high $N_{H}$) AGNs. Using this threshold we find that $96\% \pm 3$\% ($53 \pm 1$) of the BL AGNs are unobscured,
whilst $54\% \pm 10$\% ($13 \pm 2$) of the NL AGNs are unobscured according to this definition. 
Of the GALs, $91\% \pm 6$\% ($21 \pm 1$) are characterized by HR$<-0.2$ and therefore show a soft X-ray spectrum 
(notice that \citep{hasinger08} decided to classify objects without a spectroscopic AGN classification but with  
HR$<-0.2$ as a BL AGN, and therefore, under this scheme, the great majority of our GAL objects should be reclassified as AGN).

Following our definition of variability consistent with being produced by the accretion disk of a super-massive black hole
(V \textgreater 1.3, $\gamma$ \textgreater $0.025$, and $A$ \textgreater $0.1$) we find that $98\% \pm 2$\% ($43 \pm 1$) 
of the BL AGNs that meet these criteria are unobscured, $100$\% ($3$) of the NL AGNs that meet these criteria are obscured, 
and assuming that the X-ray emission of the GALs that meet these criteria is produced by low luminosity AGN $100$\% ($2$) of them
are unobscured. As it is expected beforehand the majority of the variable BL AGNs are unobscured. However, it is interesting that, although
in the NL AGNs there is a large spread in the HRs and that $54$\% of them are unobscured in the best observed sample, all the 
variable NL AGNs are obscured (see Figure \ref{V_vs_midir_HR_fig}). This result is also confirmed by the fact
that the variable NL AGNs with the larger variability amplitude ($\sigma_{rms}$ \textgreater $0.05$ mag) are among obscured objects.
The most likely interpretation is that a fraction of the accretion disk emission is leaking through the obscuring material.

\subsection{The XMM-COSMOS Variable Objects}

\begin{deluxetable*} {lccccccccc}
\tablecolumns{10}
\tablenum{6}
\tablewidth{0pc}
\tablecaption{Properties of Some of the Most Variable XMM-COSMOS Objects.\label{PropExLC_tab}}

\tablehead{
\colhead{Name} &
\colhead{Class} &
\colhead{$z$} &
\colhead{$r$ mag} &
\colhead{$P_{var}$\tablenotemark{a}} &
\colhead{$\sigma_{rms}$} &
\colhead{$[3.6]-[8.0]$} &
\colhead{Host/AGN} &
\colhead{HR}  &
\colhead{Obscured/} \\
\colhead{} &
\colhead{} &
\colhead{} &
\colhead{} &
\colhead{} &
\colhead{} &
\colhead{AB color} &
\colhead{dominated\tablenotemark{b}} &
\colhead{} &
\colhead{Unobscured\tablenotemark{c}} 
}
\startdata

XID~59    & BL  & $1.920$ & $20.50$ & \textgreater $0.999$ & $0.258$ & $1.41$ & AGN & $-0.51$ & unobscured \\
XID~5607  & BL  & $1.359$ & $20.52$ & \textgreater $0.999$ & $0.197$ & $1.70$ & AGN & $-1.00$ & unobscured \\
XID~5544  & BL  & $1.889$ & $20.21$ & \textgreater $0.999$ & $0.108$ & $1.54$ & AGN & $-0.55$ & unobscured \\
XID~293   & NL  & $0.445$ & $20.32$ & \textgreater $0.999$ & $0.137$ & $0.93$ & AGN & $-0.02$ & obscured \\
XID~5047  & NL  & $0.252$ & $20.10$ & \textgreater $0.999$ & $0.068$ & $0.25$ & host & $1.00$ & obscured \\
XID~1429  & NL  & $0.356$ & $19.77$ & $0.999$ & $0.025$ & $0.87$ & AGN & $-0.20$ & obscured \\  
XID~10569 & GAL & $0.348$ & $19.91$ & \textgreater $0.999$ & $0.067$ & $-1.15$ & host & $-1.00$ & unobscured \\
XID~60406 & GAL & $0.166$ & $19.77$ & $0.978$ & $0.038$ & $-1.54$ & host & $-1.00$ & unobscured \\

\enddata
\tablenotetext{a}{We defined $P_{var} = 1 - P(\chi^{2})$ (see Section \ref{variability}), and is interpreted as
the probability of an object to be variable based on its $\chi^{2}$ value and the number of observations.}
\tablenotetext{b}{We used the division between AGN or host dominated sources of \citet{brusa10}, that is 
based on the mid-IR color ($[3.6] - [8.0]$) of the source.}
\tablenotetext{c}{We follow the convention that sources with HR \textless $-0.2$ are unobscured.}
\end{deluxetable*}

\begin{figure*}
\plottwo{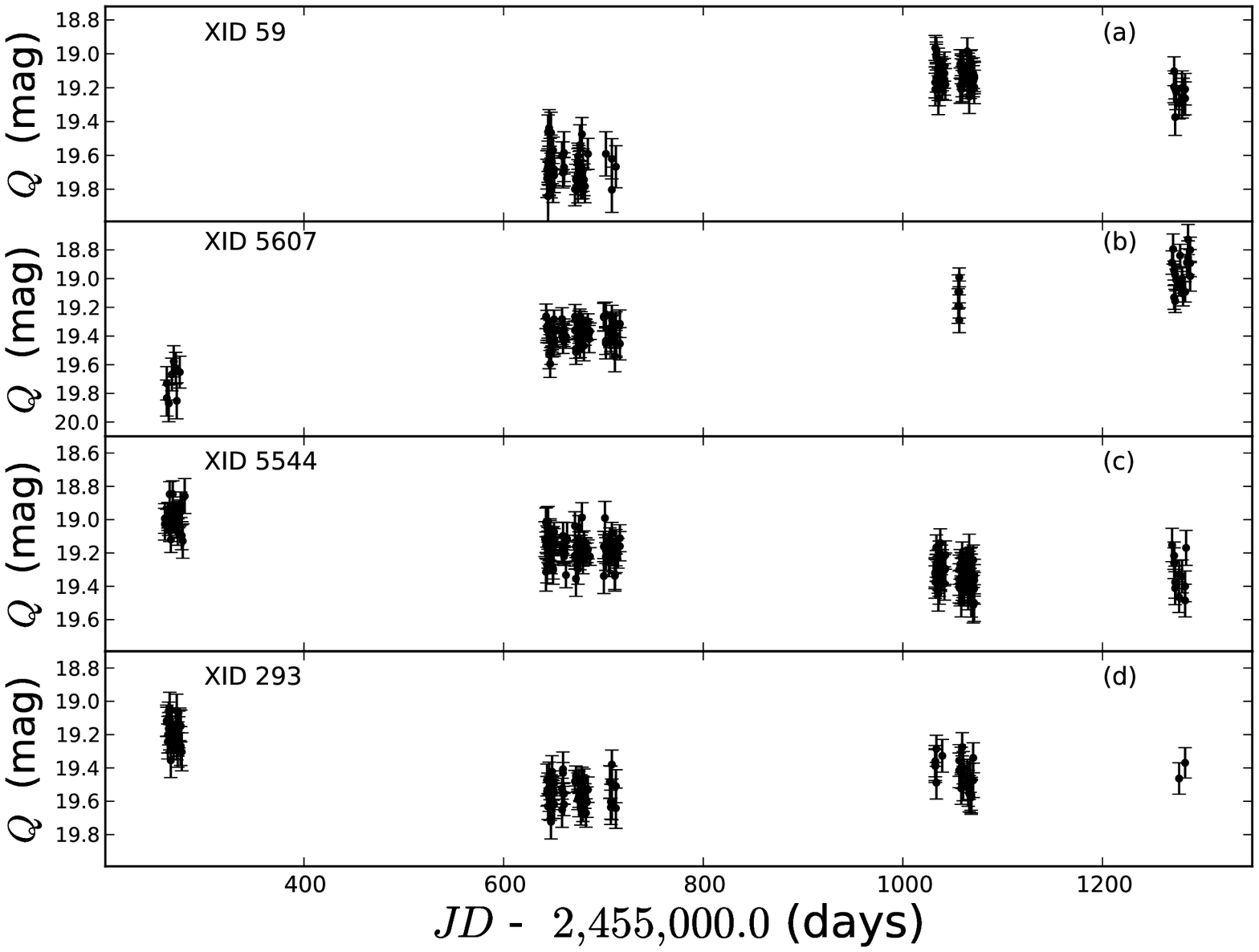}{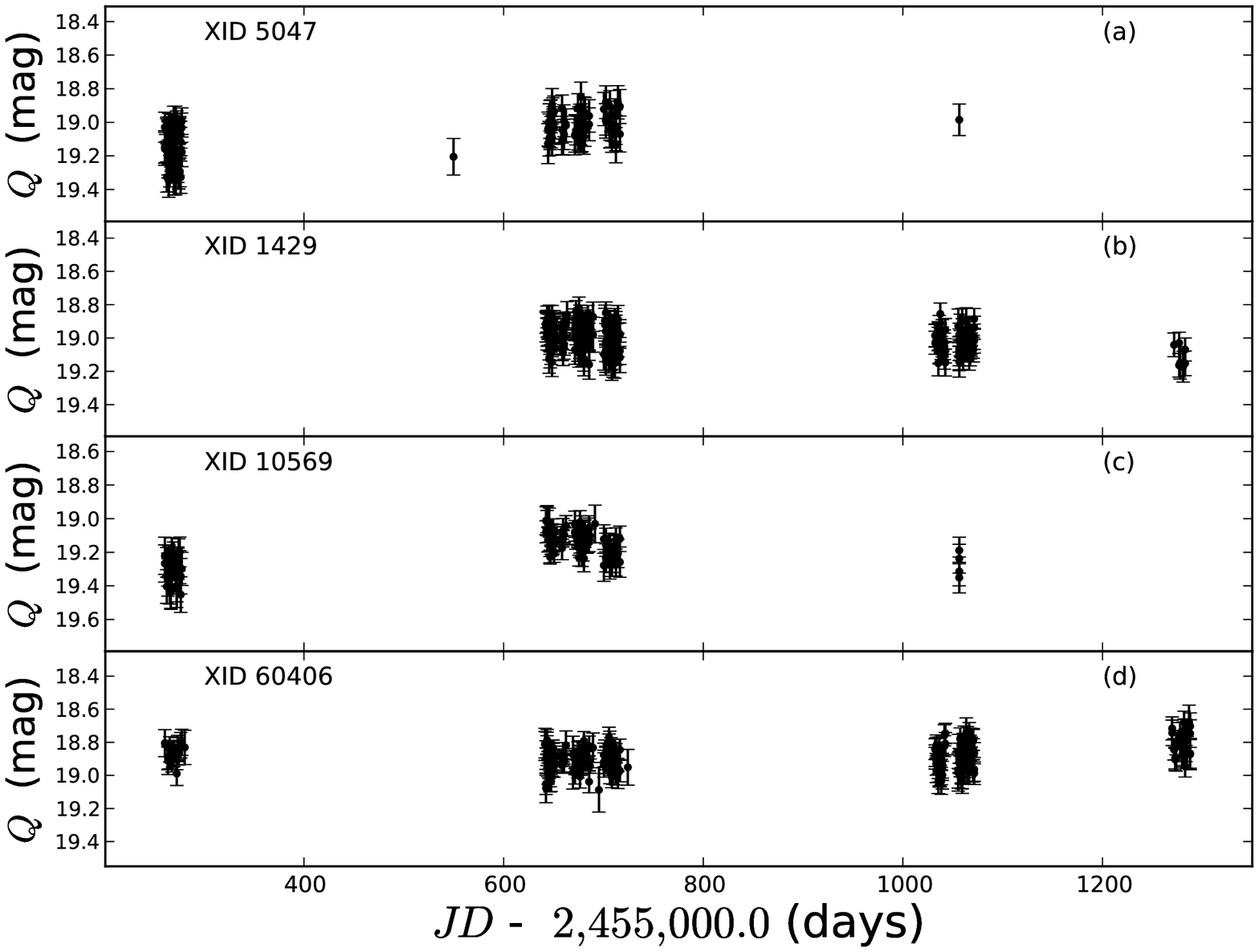}
\caption{$Q$-band light curves of XMM-COSMOS variable objects. In the top-right of each panel is the XMM-COSMOS identifier number for each object.}
\label{ex_var_fig}
\end{figure*}

In Figure \ref{ex_var_fig} we show eight light curves of the most
optically variable sources among the XMM-COSMOS objects whose observed
variability is consistent with being produced by an AGN, including the 3
NL AGNs and the 2 GALs that are variables and their structure function
parameters are consistent with variability produced by an accretion
disk as in the case of BL AGNs. In the top-right of each panel
the XMM-COSMOS identifier number for each object is shown, while the
properties of the objects are summarized in Table
\ref{PropExLC_tab}. All BL AGNs show clear variability with large
amplitudes ($\sigma_{rms}$ \textgreater $0.05$ mag). The NL AGNs show
mid-IR colors close to the division between AGN and host dominated
objects and all have HR values consistent with gas obscuration. A likely
interpretation for these objects is disk emission leaking through the
obscuring material.

Previous studies have found that between 1\%-4\%\ of field galaxies
show variability that could be associated with low-luminosity AGNs
\citep{sarajedini00}. The effects of the emission produced by the
accreting black hole on the spectrum and the images of the galaxies
containing a low-luminosity AGNs are rarely detectable. For example,
broad emission lines are not present in the galaxy spectrum even if
the AGN is not obscured as the AGN emission is heavily diluted by the
host galaxy light. If there is an underlying low-luminosity AGN a
careful subtraction of two images of the galaxy may uncover the
presence of a variable AGN \citep{sarajedini00}. Since in our
survey we do not subtract a reference image to the rest of the images,
a reliable identification of these low-luminosity AGNs is
difficult. Given that our sample of galaxies is X-ray selected and
some of them show narrow-emission lines, the probability of finding a
low-luminosity AGN is higher than for a sample of normal field
galaxies. Among our sample of GALs we found that 22\%\ of them show
variability and 8.7\%\ show variability consistent with an accretion
disk. In particular, XID 10569 is clearly variable (rms $> 0.05$ mag)
and consistent with the variation expected for an AGN.

\section{DISCUSSION AND CONCLUSIONS}
\label{summary}

In this paper we present the characterization of the QUEST-La Silla AGN variability survey. The survey 
is a novel effort to obtain highly sampled optical light curves in well studied extragalactic 
fields, making use of the QUEST camera at the ESO-Schmidt telescope at La Silla. Our observing
strategy will provide good quality light curves for many sources that already have a large data set of 
multi-wavelength observations, and will increase the number of AGN candidates in the areas of the survey 
where a AGN classification is missing or is based only on a color-color selection \citep{schmidt10, butler11, palanque11, graham14}.
Additionally, the survey is producing highly sampled light curves for all classes of interesting transients
\citep[see e.g.,][]{rabinowitz12, zinn14, cartier15}.

In this paper, we showed that we are obtaining good astrometry, with an internal precision of $\sim 0\farcs1$ and with
an overall accuracy of $0\farcs2$ compared to SDSS. In the future, the QSOs lying within our observing fields may
be used as references to increase the internal astrometric precision of the survey.

We defined the $Q$-band as the union of the $g$ and $r$ bands, and we showed that calibrating
the $Q$-band as an equivalent to the $(g+r)_{SDSS}$ photometric system yields a photometric
dispersion of $\sim 0.05$ magnitudes. Furthermore, we quote, conservatively, a systematic error
in our zero-point of $0.05$ magnitudes. 

We found that the use of a linear model to fit the zero-point as function of the instrumental magnitude $q$
produces a better calibration of our data, increasing significantly the number of non-variable stars, and
reducing the skewness in the $Q-(g+r)_{SDSS}$ residuals as function of $Q$.

We demonstrated that we are obtaining good photometry in the range $14.0 \lesssim Q \lesssim 19.5$. 
Since we are observing a minimum of two images per tile in a given night, we expect to produce light curves 
using nightly stacked images. Our preliminary results indicate that in the near future we will produce good quality 
light curves with extraordinary cadence of observations to a magnitude limit of $Q \sim 20.5$ (roughly $r \sim 21$),
or deeper. 

As a way to explore the quality of the data collected by our survey, we studied the optical variability of X-ray
XMM-COSMOS sources of \citet{brusa10}. In this study, we found that the QUEST-La Silla AGN variability survey 
is $\sim 75$\% to $80$\% complete in the XMM-COSMOS field to magnitude limit of $r \sim 20$, and $\sim 67$\%
complete to a magnitude of $r \sim 21$. Therefore, the survey loses roughly $20$\% of the objects, possibly
due to large defective areas, permanently or randomly off detectors, and gaps between detectors. 

Based on the variability index $V$ (see Section \ref{variability}), we found that $80\% \pm 5$\% of the BL 
objects are classified as variable objects, while  $21\% \pm 8$\% of the NL objects and the $22\% \pm 9$\% of 
the GAL objects are classified as variable objects.

We studied the relation between $\sigma_{rms}$ and the $[3.6] - [8.0]$ mid-IR color, and found that objects that are redder in the 
mid-IR (possibly AGN-dominated) tend to show larger $\sigma_{rms}$. A possible interpretation of this result is that the emission
of the host galaxy dilutes AGN variability in objects that are more host galaxy dominated. We also found that overall
BL objects have larger variation amplitudes (e.g., higher $\sigma_{rms}$) than NL, and GAL objects. However,
a significant fraction ($\sim 33$\%) of NL variable objects show significant variability.

We studied the variability as function of the time elapsed between observations ($\tau$) in the {\it best observed\/} 
sample using the ``normalized'' structure function ($SF_{norm}$),  i.e. after correction for measurement errors. We
parametrized $SF_{norm}(\tau)$ as a power-law ($A \tau^{\gamma}$), and
found that BL AGN tend to be located in a region with large amplitudes ($A > 0.1$)
and power-law exponents ($\gamma > 0.025$) compared to the bulk of NL AGNs and GALs. We also demonstrated that the 
variability in variable NL AGNs and GALs with $A > 0.1$ and $\gamma > 0.025$ seems to be the result of an
accreting super-massive black hole as in the case of BL AGNs and not an artifact of the observations. 
We found that 74.5\% of BL AGNs, 12.4\% of NL AGNs and 8.7\% of GALs are variables and have $A > 0.1$ and $\gamma > 0.025$.

We confirm that the majority of the BL variable objects have mid-IR colors ($[3.6] - [8.0]$) consistent with being AGN-dominated,
and hardness ratios (HRs) consistent with unobscured X-ray sources. On the other hand, we found that although $54$\% $\pm 10$\%
of the NL objects in the {\it best observed\/} sample are unobscured according to the HR division, 100\% of the NL AGNs
showing variability consistent with being powered by a super-massive black hole ($V > 1.3$, $\gamma > 0.025$ and $A > 0.1$) are obscured. 
This is an interesting result that requires confirmation from larger samples with high quality light curves. 
The time span of our QUEST-La Silla light curves will continue to increase until 2016, so we expect to revisit this
result in a future paper.

The $[3.6]-[8.0]$ mid-IR color distribution of NL objects shows a large spread, with a group of NL
objects showing very blue colors, and a second group with redder colors peaking close to the division between AGN dominated
and host dominated sources of \citet{brusa10}, but located on the blue side of the BL distribution (see Figure \ref{V_vs_midir_HR_fig}).

Although the NL AGN sample is small, we noticed that the objects with larger excess variance and with $\gamma > 0.025$ and $A > 0.1$ 
are those with relatively red mid-IR colors and larger values of HR (see Figure \ref{midir_vs_sigma_fig}). Thus, a possible interpretation
of variable NL objects is that they are obscured sources, in which the optical variability could be the result of leaked or scattered disk
emission. However, an increase of the sample size is definitely required to have conclusive interpretations.

GALs show a large spread in their $[3.6] - [8.0]$ color distribution. However, all variable GAL sources have
blue mid-IR colors, and thus are consistent with being host-dominated objects. Additionally, assuming that variable GALs with 
$\gamma > 0.025$ and $A > 0.1$ contain an AGN, we find that these are unobscured X-ray sources according
to the HR definition. Consequently, a plausible interpretation for the variable GAL objects having $\gamma > 0.025$
and $A > 0.1$ is that they are low-luminosity AGNs.

\acknowledgments

\noindent
We thank to the anonymous referee for his careful review that helped to improve this manuscript.
R.C. acknowledges support by CONICYT through ``Programa Nacional de Becas de Postgrado'' grant D-2108082, by the Yale-Chile fellowship in astrophysics, 
by Centro de Astrof\'\i sica FONDAP 15010003, Center of Excellence in Astrophysics and Associated Technologies (PFB 06), by the 
``Millennium Center for Supernova Science'' through grant P10-064-F,  and the ``Millennium Institute of Astrophysics (MAS)'' 
through grant IC120009 of the ``Programa Iniciativa Cient\'ifica Milenio del Ministerio de Econom\'ia, Fomento y Turismo de Chile''. 
P.~L. acknowledge support from FONDECYT through grant 1120328.\\
This research has made use of the VizieR catalogue access tool, CDS, Strasbourg, France.
Funding for the SDSS and SDSS-II has been provided by the Alfred P. Sloan Foundation, the Participating Institutions,
the National Science Foundation, the U.S. Department of Energy, the National Aeronautics and Space Administration,
the Japanese Monbukagakusho, the Max Planck Society, and the Higher Education Funding Council for England.
The SDSS Web Site is http://www.sdss.org/.
The SDSS is managed by the Astrophysical Research Consortium for the Participating Institutions. The Participating
Institutions are the American Museum of Natural History, Astrophysical Institute Potsdam, University of Basel,
University of Cambridge, Case Western Reserve University, University of Chicago, Drexel University, Fermilab,
the Institute for Advanced Study, the Japan Participation Group, Johns Hopkins University, the Joint Institute for
Nuclear Astrophysics, the Kavli Institute for Particle Astrophysics and Cosmology, the Korean Scientist Group,
the Chinese Academy of Sciences (LAMOST), Los Alamos National Laboratory, the Max-Planck-Institute for Astronomy (MPIA),
the Max-Planck-Institute for Astrophysics (MPA), New Mexico State University, Ohio State University, University of Pittsburgh,
University of Portsmouth, Princeton University, the United States Naval Observatory, and the University of Washington.
Based on observations obtained with XMM-Newton, an ESA science mission with instruments and contributions directly
funded by ESA Member States and NASA.


\begin{thebibliography}{}

\bibitem[Abazajian et~al.(2009)]{abazajian09} Abazajian, K.~N., Adelman-McCarthy, J.~K., Ag\"ueros, M.~A., et~al. 2009, \apjs, 182, 543

\bibitem[Abbott et~al.(2005)]{abbott05} Abbott, T., Aldering, G., Annis, J., et~al. (The Dark Energy Survey Collaboration) 2005, (arXiv:astro-ph/0510346)

\bibitem[Adelman-McCarthy et~al.(2006)]{adelman-mccarthy06} Adelman-McCarthy, J.~K., Ag\"ueros, M.~A., Allam, S.~S., et~al. 2006, \apjs, 162, 38

\bibitem[Ai et~al.(2010)]{ai10} Ai, Y.~L., Yuan, W., Zhou, H.~Y., et~al. 2010, \apj, 716, L31

\bibitem[Ar\'evalo et~al.(2008)]{arevalo08} Ar\'evalo, P., Uttley, P., Kaspi, S., et~al. 2008, \mnras, 389, 1479

\bibitem[Ar\'evalo et~al.(2009)]{arevalo09} Ar\'evalo, P., Uttley, P., Lira, P., et~al. 2009, \mnras, 397, 2004

\bibitem[Assef et~al.(2013)]{assef13} Assef, R.~J., Stern, D., Kochanek, C.~S., et~al. 2013, \apj, 772, 26
  
\bibitem[Baltay et~al.(2007)]{baltay07} Baltay, C., Rabinowitz, D., Andrews, P., et~al. 2007, \pasp, 119, 1278


\bibitem[Bertin \& Arnouts(1996)]{bertin96} Bertin, E. \& Arnouts, S. 1996, \aap 117, 393

\bibitem[Bertin(2006)]{bertin06} Bertin, E. 2006, in Astronomical Data Analysis Software and Systems XV, ASP Conf. Series, 351, 112

\bibitem[Brandt \& Hasinger(2005)]{brandt05} Brandt, W. N. \& Hasinger, G., 2005, ARA\&A, 43, 827

\bibitem[Brusa et~al.(2010)]{brusa10} Brusa, M., Civano, F., Comastri, A., et~al. 2010, \apj, 716, 348

\bibitem[Butler \& Bloom(2011)]{butler11} Butler, N.~R. \& Bloom, J.~S. 2011, \aj, 141, 93 




\bibitem[Capak et~al.(2007)] {capak07} Capak, P., Aussel, H., Ajiki, M., et~al. 2007, \apjs, 172, 99

\bibitem[Cartier et~al.(2015)]{cartier15} Cartier, R., Lira, Sanchez, P., Coppi, P., et~al. 2015, (in preparation)

\bibitem[Collier \& Peterson(2001)] {collier01} Collier, A. \& Peterson, B.~M., 2001, \apj, 555, 775

\bibitem[Cristiani et~al.(1997)]{cristiani97} Cristiani, S., Trentini, S., La Franca, F. \& Andreani, P. 1997, \aap, 321, 123

\bibitem[Drake et~al.(2009)]{drake09} Drake, A.~J., Djorgovski, S.~G., Mahabal, A., et~al. 2009, \apj, 696, 870

\bibitem[Elvis et~al.(2009)]{elvis09} Elvis, M., Civano, F., Vignali, C., et~al., 2009, \apjs, 184, 158

\bibitem[Fan (1999)]{fan99} Fan, X. 1999, \aj, 117, 2528

\bibitem[Frieman et~al.(2008)]{friedman08} Frieman, J., Bassett, B., Becker, A., et~al. 2008, \aj, 135, 338

\bibitem[Fukugita et~al.(1996)]{fukugita96} Fukugita, M., Ichikawa, T., Gunn, J.~E., et~al. 1996, \aj, 111, 1748


\bibitem[Graham et~al.(2014)]{graham14} Graham, M.~J., Djorgovski, S.~G., Drake, A.~J., et~al., 2014, \mnras, 439, 703

\bibitem[Gierli\'nski et~al.(2008)] {gierlinski08} Gierli\'nski, M., Niko{\/l}ajuk, M., Czerny, B. 2008, \mnras, 383, 741


\bibitem[Hasinger(2008)]{hasinger08} Hasinger, G. 2008, \aap, 490, 905

\bibitem[Hamuy et~al.(1994)]{hamuy94} Hamuy, M., Suntzeff, N.~B., Heathcote, S.~R., et~al. 1994, \pasp, 106, 566

\bibitem[Hamuy et~al.(1992)]{hamuy92} Hamuy, M., Walker, A.~R., Suntzeff, N.~B., et~al. 1992, \pasp, 104, 533

\bibitem[Hodapp et~al.(2004)]{hodapp04} Hodapp, K.~W., Kaiser, N., Aussel, H., et al., 2004, AN, 325, 63

\bibitem[Hook et~al.(1994)]{hook94} Hook, I.~M., McMahon, R.~G., Boyle, B.~J. \& Irwin, M.~J. 1994, \mnras, 268, 305

\bibitem[Ilbert et~al.(2009)]{ilbert09} Ilbert, O., Capak, P., Salvato, M., et~al. 2009, \apj, 690, 1236

\bibitem[Ivezi\'c et~al.(2004)]{ivezic04} Ivezi\'c, Z., Smith, J.~A., Miknaitis, G., et~al. 2004, \aj, 134, 937

\bibitem[Ivezi\'c et~al.(2008)]{ivezic08} Ivezi\'c, Z., Tyson, J.~A., Acosta, E., et~al. (LSST Collaboration), [arXiv:0805.2366]

\bibitem[Jarvis et~al.(2013)]{jarvis13} Jarvis, M.~J., Bonfield, D.~G., Bruce, V.~A., et~al., 2013, \mnras, 428, 1281

\bibitem[Kauffmann et~al.(2003)]{kauffmann03} Kauffmann, G., Heckman, T.~M., Tremonti, C., et~al. 2003, \mnras, 346, 1055 

\bibitem[Kelly et~al.(2009)]{kelly09} Kelly, B.~C., Bechtold, J., \& Siemiginowska, A., 2009, \apj, 698, 895

\bibitem[Kelly et~al.(2013)]{kelly13} Kelly, B.~C., Treu, T., Malkan, M., Pancoast, A., \& Woo, J. 2013, \apj, 778, 187 


\bibitem[Kim et~al.(2012)]{kim12} Kim, D.~-W., Protopapas, P., Trichas, M., et~al. 2012, \apj, 747, 107

\bibitem[Koz{\l}owski \& Kochanek(2009)]{kozlowski09} Koz{\l}owski, S. \& Kochanek, C.~S. 2009, \apj, 710, 508

\bibitem[Koz{\l}owski et~al.(2010)]{kozlowski10} Koz{\l}owski, S., Kochanek, C.~S., Udalski, A., et~al. 2010, \apj, 708, 927


\bibitem[Lacy et~al.(2004)]{lacy04} Lacy, M., Storrie-Lombardi, L.~J., Sajina, A., et~al. 2004, \apjs, 154, 166

\bibitem[Lang et~al.(2010)]{lang10} Lang, D., Hogg, D.~W., Mierle, K., Blanton, M., \& Roweis, S. 2010, \apj, 139, 1782

\bibitem[Lanzuisi et~al.(2014)]{lanzuisi14} Lanzuisi, G., Ponti, G., Salvato, M., et~al. 2014, \apj, 781, 105

\bibitem[Law et~al.(2009)]{law09} Law, N.~M., Kulkarni, S.~R., Dekany, R.~G., et~al. 2009, \pasp, 121, 1395

\bibitem[Le~Floc'h et~al.(2009)]{lefloch09} Le~Floc'h, E.,  Aussel, H., Ilbert, O., et~al. 2009, \apj, 703, 222


\bibitem[Lilly et~al.(2007)]{lilly07} Lilly, S.~J., Le~F\`evre, O., Renzini, A., et~al. 2007, \apjs, 172, 70 

\bibitem[Lilly et~al.(2009)]{lilly09} Lilly, S.~J., Le~Brun, V., Maier, C., et~al. 2009, \apjs, 184, 218

\bibitem[Lira et~al.(2011)]{lira11} Lira, P., Ar\'evalo, P., Uttley, P., McHardy, I., Breedt, E., 2011, \mnras, 415, 1209

\bibitem[Lira et~al.(2013)]{lira13} Lira, P., Videla, L., Wu, Y., et~al. 2013, \apj, 764, 159

\bibitem[Lira et~al.(2015)]{lira15} Lira, P., Ar\'evalo, P., Uttley, P., McHardy, I.~M.~M., Videla, L., 2015, \mnras, submitted


\bibitem[MacLeod et~al.(2010)]{macleod10} MacLeod, C.~L., Ivezi\'c, Z., Kochanek, C.~S., et~al. 2010, \apj, 721, 1014

\bibitem[MacLeod et~al.(2011)]{macleod11} MacLeod, C.~L., Books, K., Ivezi\'c, Z., et~al. 2011, \apj, 728, 26

\bibitem[Mainieri et~al.(2007)]{mainieri07} Mainieri, V., Hasinger, G., Cappelluti, N., et~al. 2007, \apjs, 172, 368


\bibitem[McCracken et~al.(2010)]{mccracken10} McCracken, H.~J., Capak, P., Salvato, M., et~al. 2010, \apj, 708, 202

\bibitem[McCracken et~al.(2012)]{mccracken12} McCracken, H.~J., Milvang-Jensen, B., Dunlop, J., et~al. 2012, \aap, 544, 156

\bibitem[McHardy(1988)]{mchardy88} McHardy, I.~M. 1988, Mem. Soc. Astron. Italiana, 59, 239

\bibitem[McHardy et~al.(2006)]{mchardy06} McHardy, I.~M., Koerding, E., Knigge, C., Uttley, P. \& Fender, R.~P. 2006, \nat, 444, 730

\bibitem[McHardy(2013)]{mchardy13} McHardy, I.~M. 2013, \mnras, 430, L49

\bibitem[McLaughlin et~al.(1996)]{mclaughlin96} McLaughlin, M.~A., Mattox, J.~R., Cordes, J.~M., \& Thompson, D.~J. 1996, \apj, 473, 763

\bibitem[Meusinger et~al.(2013)]{meusinger13} Meusinger, H. \& Weiss, V. 2013, \aap, 506, 104

\bibitem[Morganson et~al.(2014)]{morganson14} Morganson, E., Burgett, W.~.S., Chambers, K.~C., et~al. 2014, \apj, 784, 92

\bibitem[Monet et~al.(2003)]{monet03} Monet, D.~G., Levine, S.~E., Canzian, B., et~al. 2003, \aj, 125, 984

\bibitem[Mor \& Netzer(2012)]{mor12} Mor, R. \& Netzer, H. 2012, \mnras, 420, 526


\bibitem[Palanque--Delabrouille et~al.(2011)]{palanque11} Palanque--Delabrouille, N., Yeche, Ch., Myers, A.~D., et~al. 2011, \aap, 530, 122

\bibitem[Paolillo et~al.(2004)]{paolillo04} Paolillo, M., Schreier, E.~J., Giacconi, R., Koekemoer, A.~M., \& Grogin, N.~A. 2004, \apj, 611, 93


\bibitem[Polletta et~al.(2007)]{polletta07} Polletta, M., Tajer, M., Maraschi, L., et~al. 2007, \apj, 663, 81

\bibitem[Prescott et~al.(2006)]{prescott06} Prescott, M.~K.~M., Impey, C.~D., Cool, R.~J., \& Scoville, N.~Z. 2006, \apj, 644, 100

\bibitem[Rau et~al.(2009)]{rau09} Rau, A., Kulkarni, S.~R., Law, N.~M., et~al. 2009, \pasp, 121, 1334

\bibitem[Rabinowitz et~al.(2012)]{rabinowitz12} Rabinowitz, D., Schwamb, W.~E., Hadjiyska, E., \& Tourtellotte, S. 2012, \aj, 144, 140

\bibitem[Richards et~al.(2002)]{richards02} Richards, G.~T., Fan, X., Newberg, H.~J., et~al. 2002, \aj, 123, 2945

\bibitem[Richards et~al.(2009)]{richards09} Richards, G.~T., Myers, A.~D., Gray, A.~G., et~al. 2009, \apjs, 180, 67

\bibitem[S\'anchez et~al.(2015)]{sanchez15} S\'anchez, P., Lira, P., Cartier, R., et~al. 2015 (in preparation)

\bibitem[Sanders et~al.(2007)]{sanders07} Sanders, D.~B., Salvato, M., Aussel, H., et~al. 2007, \apjs, 172, 86


\bibitem[Sarajedini et~al.(2000)]{sarajedini00} Sarajedini, V.~L., Gilliland, R.~L., Phillips, M.~M. 2000, \aj, 120, 2825

\bibitem[Schlegel et~al.(1998)]{schlegel98} Schlegel, D.~J., Finkbeiner, D.~P. \& Davis, M. 1998, \apj, 500, 525

\bibitem[Schmidt \& Green(1983)]{schmidt83} Schmidt, M. \& Green, R.~F. 1983, \apj, 269, 352

\bibitem[Schmidt et~al.(2010)]{schmidt10} Schmidt, K.~B., Marshall, P.~J., Rix, H.-W., et al. 2010, \apj, 714, 1194

\bibitem[Sesar et~al.(2007)]{sesar07} Sesar, B., Ivezi\'c, Z., Lupton, R.~H., et al. 2007, \aj, 134, 2236

\bibitem[Shemmer et~al.(2014)]{shemmer14} Shemmer, O., Brandt, W.~N., Paolillo, M., et~al. 2014, \apj, 783, 116

\bibitem[Stern et~al.(2005)]{stern05} Stern, D., Eisenhardt, P., Gorjian, V., et~al. 2005, \apj, 631, 163

\bibitem[Stetson(1987)]{stetson87} Stetson, P.~B. 1987, PASP, 99, 191

\bibitem[Tonry et~al.(2012)]{tonry12} Tonry, J.~L., Stubbs, C.~W., Lykke, K.~R., et~al. 2012, \apj, 750, 99

\bibitem[Trevese et~al.(1994)]{trevese94} Trevese, D., Kron, R.~G., Majewski, S.~R., Bershady, M.~A. \& Koo, D.~C. 1994, \apj, 433, 494

\bibitem[Trump et~al.(2007)]{trump07} Trump, J.~R., Impey, C.~D., McCarthy, P.~J., et~al. 2007, \apjs, 172, 383

\bibitem[Trump et~al.(2009)]{trump09} Trump, J.~R., Impey, C.~D., Elvis, M., et~al. 2009, \apj, 696, 1195

\bibitem[Uomoto et~al.(1976)]{uomoto76} Uomoto, A.~K., Wills, B.~J. \& Wills, D. 1976, \aj, 81, 905

\bibitem[Vanden Berk et~al.(2001)]{vandenberk01} Vanden Berk, D.~A., Richards, G.~T., Bauer, A., et~al. 2001, \aj, 122, 549 

\bibitem[Vanden Berk et~al.(2004)]{vandenberk04} Vanden Berk, D.~A., Wilhite, B.~C., Kron, R.~G., et~al. 2004, \apj, 601, 692 


\bibitem[Wilhite et~al.(2008)]{wilhite08} Wilhite, B.~C., Brunner, R.~J., Grier, C.~J., Schneider, D.~P., \& Vanden Berk, D.~E. 2008, \mnras, 383, 1232 

\bibitem[Wold et~al.(2007)]{wold07} Wold, M., Brotheron, M.~S., \& Shang, Z. 2007, \mnras, 375, 989 

\bibitem[York et~al.(2000)]{york00} York, D.~G., et~al. 2000, \aj, 120, 1579

\bibitem[Zinn et~al.(2014)]{zinn14} Zinn, R., Horowitz, B., Vivas, A.~K., et~al. 2014, \apj, 781, 22

\bibitem[Zuo et~al.(2012)]{zuo12} Zuo, W., Wu, X.-B., Liu, Y.-Q., \& Jiao, C.-L. 2012, \apj, 758, 104

\end{thebibliography}
\end{document}